%% file: EXO-22-019_temp.tex
\pdfoutput=1
\documentclass[11pt,twoside,a4paper,cmspaper,final,collab]{cms-tdr}
\def\svnVersion{e7fcdb3}\def\svnDate{2024/07/02}\def\cmsCernNoTag{CERN-EP-2024-053}\def\cmsCernDate{\today}\def\cmsMessage{Published in the Journal of High Energy Physics as \href{http://dx.doi.org/10.1007/JHEP06(2024)183}{\doi{10.1007/JHEP06(2024)183}.}}
\begin{document}\cmsNoteHeader{EXO-22-019}

\newcommand{\ctau}{\ensuremath{c\PGt_{\PN}}\xspace}
\newcommand{\tauN}{\ensuremath{\PGt_{\PN}}\xspace}
\newcommand{\mN}{\ensuremath{m_\PN}\xspace}
\newcommand{\Bu}{\ensuremath{\mathrm{B}_{\PQu}}\xspace}
\newcommand{\Bd}{\ensuremath{\mathrm{B}_{\PQd}}\xspace}
\newcommand{\Bs}{\ensuremath{\mathrm{B}_{\PQs}}\xspace}
\newcommand{\Bc}{\ensuremath{\mathrm{B}_{\PQc}}\xspace}
\newcommand{\Bq}{\ensuremath{\mathrm{B}_{\PQq}}\xspace}
\newcommand{\main}{(\Bu, \Bd, \Bs)\xspace}
\newcommand{\Vu}{\ensuremath{\abs{V_{\PGm\PN}}^2}\xspace}
\newcommand{\Ve}{\ensuremath{\abs{V_{\Pe\PN}}^2}\xspace}
\newcommand{\Vt}{\ensuremath{\abs{V_{\PGt\PN}}^2}\xspace}
\newcommand{\Vl}{\ensuremath{\abs{V_{\ell\PN}}^2}\xspace}
\newcommand{\Valpha}{\ensuremath{\abs{V_{\alpha\PN}}^2}\xspace}
\newcommand{\V}{\ensuremath{\abs{V_\PN}^2}\xspace}
\newcommand{\fc}{\ensuremath{f_{\PQc}}\xspace}
\newcommand{\fa}{\ensuremath{r_{\alpha}}\xspace}
\newcommand{\fe}{\ensuremath{r_\Pe}\xspace}
\newcommand{\fu}{\ensuremath{r_\PGm}\xspace}
\newcommand{\ft}{\ensuremath{r_\PGt}\xspace}
\newcommand{\fl}{\ensuremath{r_\ell}\xspace}
\newcommand{\Lxy}{\ensuremath{L_{xy}}\xspace}
\newcommand{\Lxysig}{\ensuremath{L_{xy}/\sigma_{L_{xy}}}\xspace}
\newcommand{\dxy}{\ensuremath{d_{xy}}\xspace}
\newcommand{\dxysig}{\ensuremath{d_{xy}/\sigma_{d_{xy}}}\xspace}
\newcommand{\dz}{\ensuremath{d_{z}}\xspace}
\newcommand{\dzsig}{\ensuremath{d_{z}/\sigma_{d_z}}\xspace}
\newcommand{\bmass}{\ensuremath{\ell_\PB\ell^\pm\PGp^\mp\mathrm{~mass}}\xspace}
\newcommand{\control}{\ensuremath{\PBpm\to\PJgy (\to\PGm^+\PGm^-)\PK^\pm}\xspace}
\newcommand{\invmass}{\ensuremath{m(\ell^{\pm}\PGp^{\mp})}\xspace}
\newcommand{\X}{\ensuremath{\PX}\xspace}

\newlength\cmsTabSkip\setlength{\cmsTabSkip}{1ex}

\cmsNoteHeader{EXO-22-019}
\title{Search for long-lived heavy neutrinos in the decays of \PB mesons produced in proton-proton collisions at \texorpdfstring{$\sqrt{s} = 13\TeV$}{sqrt(s) = 13 TeV}}

\date{\today}

\abstract{
A search for long-lived heavy neutrinos (\PN) in the decays of \PB mesons produced in proton-proton collisions at $\sqrt{s} = 13\TeV$ is presented. The data sample corresponds to an integrated luminosity of 41.6\fbinv collected in 2018 by the CMS experiment at the CERN LHC, using a dedicated data stream that enhances the number of recorded events containing \PB mesons. The search probes heavy neutrinos with masses in the range $1 < \mN < 3\GeV$ and decay lengths in the range $10^{-2} < \ctau < 10^{4}\unit{mm}$, where $\tau_\mathrm{N}$ is the \PN proper mean lifetime. Signal events are defined by the signature $\PB \to \ell_\PB \PN \mathrm{X}$; $\PN \to \ell^{\pm} \PGp^{\mp}$, where the leptons $\ell_\PB$ and $\ell$ can be either a muon or an electron, provided that at least one of them is a muon. The hadronic recoil system, X, is treated inclusively and is not reconstructed. No significant excess of events over the standard model background is observed in any of the $\ell^{\pm}\PGp^{\mp}$ invariant mass distributions. Limits at $95\%$ confidence level on the sum of the squares of the mixing amplitudes between heavy and light neutrinos, $\abs{V_\mathrm{N}}^2$, and on \ctau are obtained in different mixing scenarios for both Majorana and Dirac-like \PN particles. The most stringent upper limit $\abs{V_\mathrm{N}}^2 < 2.0\times 10^{-5}$ is obtained at $\mN=1.95\GeV$ for the Majorana case where \PN mixes exclusively with muon neutrinos. The limits on $\abs{V_\mathrm{N}}^2$ for masses $1<\mN<1.7\GeV$ are the most stringent from a collider experiment to date.
}

\hypersetup{%
pdfauthor={CMS Collaboration},%
pdftitle={Search for long-lived heavy neutrinos in the decays of B mesons produced in proton-proton collisions at sqrt(s) =13 TeV},%
pdfsubject={CMS},%
pdfkeywords={CMS, EXO, LL, HNL, BHNL}} 

\maketitle 

\section{Introduction}\label{section_introduction}

The standard model (SM) of particle physics is a highly predictive and well-tested theoretical framework that describes the electroweak and strong interactions, which together account for the known fundamental forces of nature apart from gravity. Although its predictions are in agreement with a vast number of experimental measurements, performed over a wide range of energies, the SM is unable to provide explanations for many key questions, both observational and theoretical. For example, the minimal SM does not account for the small, but nonzero neutrino masses~\cite{Bilenky:2016pep}; for the large amount of dark matter (DM) inferred from astrophysical and cosmological measurements~\cite{bertone_2010}; and for the asymmetry in the abundance of matter and antimatter in the universe~\cite{PhysRevD.50.774}. A central challenge of particle physics is therefore to develop and experimentally test new theoretical frameworks that encompass the successful predictions of the SM, but which also provide explanations for physics beyond the SM. 

An example of such an extension of the SM is the neutrino minimal standard model ($\nu$MSM)~\cite{nuMSM1, nuMSM2}. This model predicts the existence of a new type of particle, the heavy neutral lepton (HNL), also referred to as a heavy neutrino, denoted here by \PN. The dominant component of such a particle has a right-handed chirality and does not have SM gauge couplings, which is why it is often referred to as a sterile neutrino. Its interactions with SM particles arise through the admixture of an SM neutrino. In the scenario in which three heavy neutrinos are predicted, one with a mass in the keV~range and the other two with nearly degenerate masses in the GeV~range, this model can provide not only a DM candidate~\cite{DM1} but also a possible mechanism for baryogenesis~\cite{FUKUGITA198645}, as well as a possible explanation for the smallness of the masses of the SM neutrinos, through the Type~I seesaw mechanism~\cite{Minkowski:1977sc}. An overview of theoretical models that predict the existence of HNLs is presented, \eg in Ref.~\cite{Abazajian:2012ys} and references therein. 

The present search takes as a framework the phenomenology described in Ref.~\cite{bondarenko}. The underlying assumption is that the \PN states are not mass degenerate, which implies that they are produced and decay independently, or, alternatively, that their degeneracy is so close that they can be considered a single particle for all practical purposes in this search. This search can be formulated in terms of a single \PN, parametrized by its mass \mN and its mixing amplitudes to the three lepton flavour families, denoted as $V_{\Pe\PN}$, $V_{\PGm\PN}$, and $V_{\PGt\PN}$. The quantity \V is defined as $\V \equiv \Ve + \Vu + \Vt$, and the mixing ratios \fl, with $\ell=(\Pe, \PGm, \PGt)$, are defined as $\fl \equiv \Vl/\V$, with $\fe+\fu+\ft=1$ by construction. The proper mean lifetime of the heavy neutrino, \tauN, depends on \mN and \V as $\tauN \sim \abs{V_\PN}^{-2} \mN^{-5}$.
 
Searches for HNLs have been performed by several experiments using a wide range of techniques. In the mass range investigated in the present analysis, $1<\mN<3\GeV$, early searches were performed in beam-dump experiments; \eg the CHARM~\cite{CHARM}, the NuTeV~\cite{NuTeV:1999kej}, and the BEBC~\cite{Barouki_2022,WA66:1985mfx} experiments were sensitive to HNL masses up to about 2\GeV. Searches have been performed by $\Pep\Pem$~B-factory experiments, such as Belle~\cite{Belle:2013ytx} and BaBar~\cite{BaBar:2022cqj}. At the CERN LHC, limits on heavy neutral leptons have been set by the ATLAS~\cite{ATLAS2019,ATLAS2022}, CMS~\cite{CMS:trilepton,CMS:SSdilepton,CMS:hnlTau,WHNLpaper,EXO_21_013,CMS:2024ake}, and LHCb~\cite{LHCb:2014osd,LHCb2021} experiments.

This search probes long-lived \PN states that can be produced in the leptonic or semileptonic decays of B mesons, through the small admixture of SM neutrinos. A novel and key feature of the analysis is the use of a special \PQb-hadron-enriched sample of proton-proton (pp) collision data, referred to as the B-parking data sample~\cite{EXO-23-007,RKpaper}, which was collected at a centre-of-mass energy of 13\TeV by the CMS experiment. This data sample corresponds to an integrated luminosity of 41.6\fbinv and contains of order $10^{10}$ \bbbar events.

The use of \PB meson decays as the source of neutrinos provides sensitivity complementary to that of previous CMS HNL searches~\cite{CMS:trilepton,CMS:SSdilepton,CMS:hnlTau,WHNLpaper,EXO_21_013,CMS:2024ake}, which have exploited on-shell \PW~boson decays as the source of neutrinos. The \PB mesons are more abundant in LHC events than \PW~bosons and, because $m_\PB \ll m_\PW$, \PN states produced in \PB decays generally have lower momenta than those produced in \PW boson decays. 
For the long-lived HNLs considered in this search, with \ctau ranging from a few microns to a few metres, the softer momentum spectrum leads to a higher fraction of heavy neutrinos decaying within the CMS tracker volume. This volume corresponds to a distance, \Lxy, in the transverse plane between the HNL decay vertex and the luminous region of $\Lxy < 1\unit{m}$. 

Heavy neutrinos produced in \PB meson decays are kinematically constrained to have a mass $\mN \lesssim m_{\PB}$, with $m_{\PB}$ ranging from 5.27\GeV for the lighter \PBpm and \PBz mesons to $6.25\GeV$ for the heavier $\PB_\PQc^\pm$ meson. However, heavy neutrino states with masses below 1\GeV and above 3\GeV are not studied here, as this region of parameter space is well covered by other searches~\cite{EXO_21_013,Deppisch:2015qwa}. The \PB mesons considered for this search are \PBpm, \PBz, \PBzs, and $\PB_\PQc^\pm$, which we denote generically as \PBq with $\PQq = (\PQu, \PQd, \PQs, \PQc)$, respectively. Semileptonic decays of \PQb baryons are not included in the signal simulation; as a consequence of their smaller fragmentation fraction~\cite{PDGReview}, their contribution to the signal yield is expected to be significantly smaller than that from \PB mesons. The production model is discussed in more detail in Section~\ref{section_normalization}, where an explanation is also given of how the extracted signal yields are normalized using the decay \control, which has a similar topology.

\begin{figure}[tbh!]
\centering
\includegraphics[width=0.9\textwidth]{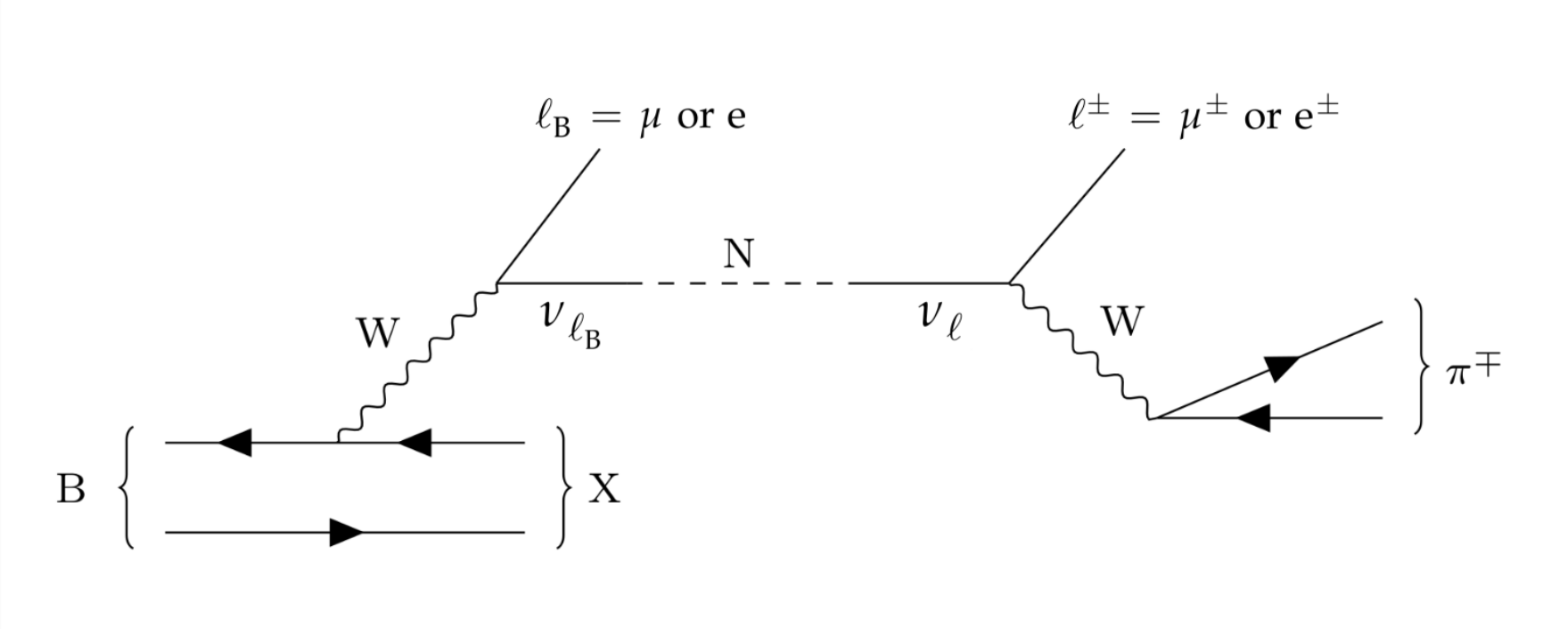}
\includegraphics[width=0.9\textwidth]{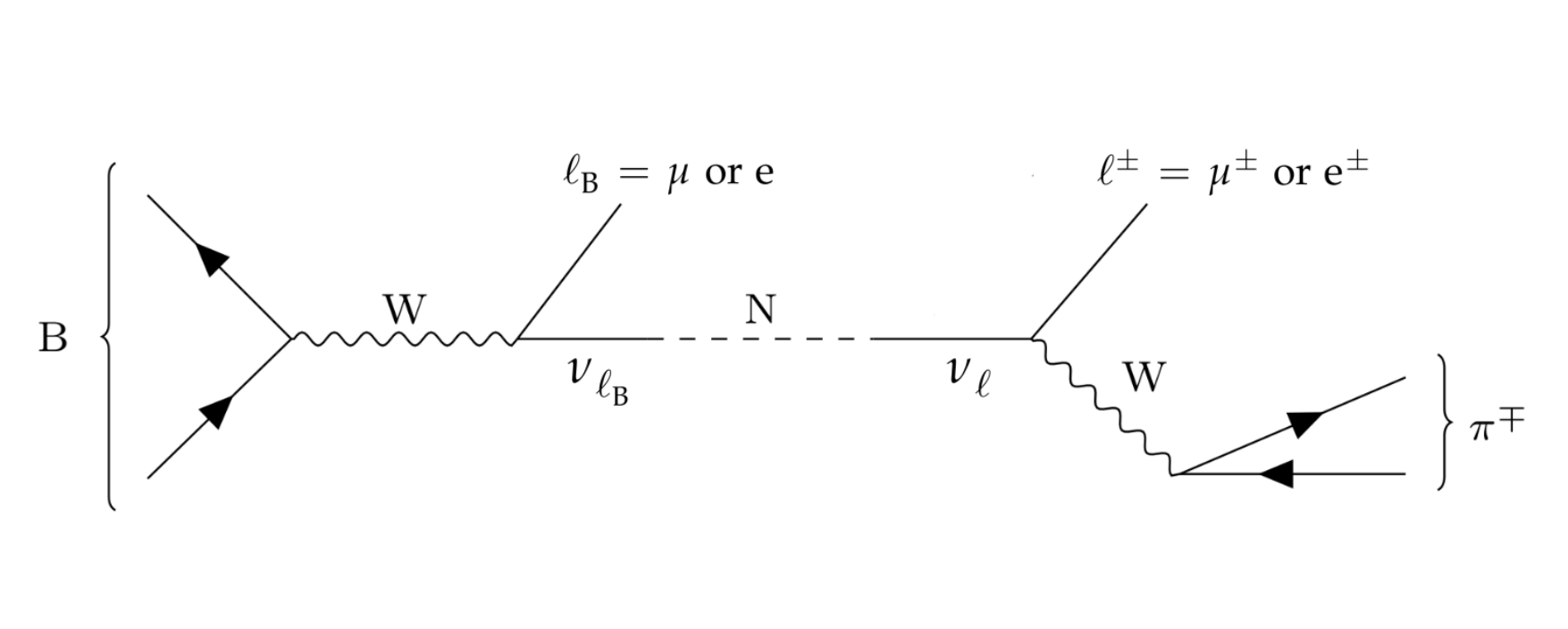}
\caption{Feynman diagrams showing the semileptonic (upper row) and leptonic (lower row) decay of a \PB meson into a lepton ($\ell_\PB$), a hadronic system (\X) in case of the semileptonic decay, and a neutrino ($\Pgn_{\ell_\PB}$), which contains a small admixture of a heavy neutrino (\PN). The \PN mass eigenstate propagates and, according to its admixture of the neutrino flavour eigenstate ($\Pgn_{\ell}$), decays weakly into a lepton $\ell^{\pm}$ and a charged pion $\PGp^{\mp}$. 
\label{figure_feynman_diagram}}
\end{figure}

The \PB meson decays considered are either semileptonic (Fig.~\ref{figure_feynman_diagram}, upper row), producing a lepton, a neutrino, and an accompanying hadronic system, \X, or leptonic (Fig.~\ref{figure_feynman_diagram}, lower row), producing a lepton and a neutrino. The contribution to the expected signal yields of the leptonic channel starts to become relevant for $\mN > 2\GeV$. To maximize the signal acceptance, the semileptonic processes are treated inclusively, without attempting to reconstruct the hadronic system \X, which can involve complex decay chains. The charged-lepton candidate from the \PB meson decay, $\ell_\PB$, can either be a muon or an electron. In a signal event, the neutrino flavour eigenstate, $\nu_{\ell_{\PB}}$, from the \PB meson decay chain contains a small admixture of an \PN mass eigenstate, which subsequently decays via its admixture of the same or another neutrino flavour eigenstate, $\nu_\ell$, into a lepton $\ell^{\pm}$, which is also required to be a muon or an electron, and a charged pion $\PGp^{\mp}$. Together, these particles form a neutral $\ell^{\pm}\PGp^{\mp}$ system, which can be reconstructed to obtain the invariant mass $\mN=\invmass$. In the region of parameter space probed here, the HNL is long lived, and its decay gives rise to a displaced vertex (DV), which is a powerful feature of the search signature. Thus, the strategy is defined by three main elements: the $\ell_\PB(\ell^{\pm}\PGp^{\mp})$ event topology, a signal peak in the invariant mass distribution of the $\ell^{\pm}\PGp^{\mp}$ system, and a DV associated with the two tracks that form this system. A strength of this analysis is that the mass of the \PN can be directly probed if a signal is present.

The search is performed under the assumption that \PN is either a Majorana or a Dirac-like neutrino~\cite{DREWES_2013}. For the case in which the \PN is assumed to be a Majorana particle, the charge of $\ell$ is not constrained by that of $\ell_\PB$, and both opposite-sign (OS) and same-sign (SS) lepton combinations, corresponding to lepton-number-conserving (LNC) and lepton-number-violating (LNV) decays, respectively, are considered. For the case in which two quasi-degenerate Majorana states form a Dirac-like state, their destructive interference cancels the LNV decays, constraining the two leptons to be OS~\cite{Hernandez:2018cgc}.

The event sample used in this analysis was recorded using single muon triggers with relatively low threshold on the muon transverse momentum, \pt, as described in Section~\ref{section_bparking}. 
The search algorithm requires that the trigger muon be part of the signal process, shown in Fig.~\ref{figure_feynman_diagram}, either as $\ell_\PB$, or as $\ell$. This muon trigger requirement leads to a classification of events according to the flavour combinations of the leptons: $(\ell_\PB,\, \ell) = (\PGm\PGm,\, \Pe\PGm,\, \PGm\Pe)$. The $\PGm\PGm$ channel is referred to as the dimuon channel, while the other two possibilities, \ie $\PGm\Pe$ and $\Pe\PGm$, are referred to as the mixed-flavour channel. 

This paper is organized as follows: the CMS detector is described in Section~\ref{section_CMS_detector}, and the data sample is discussed in Section~\ref{section_bparking}, which highlights the role of the special B-parking data stream. The simulated samples, which are only used to model signal events, are described in Section~\ref{section_simulation}. Sections~\ref{section_reconstruction} and~\ref{section_selection} describe the event reconstruction and selection, respectively. The signal normalization is discussed in Section~\ref{section_normalization}. Section~\ref{section_signal_extraction} presents the signal extraction method, which is performed using a parametric fit to the data, while Section~\ref{section_systematics} discusses the systematic uncertainties. The results are presented in Section~\ref{section_results} and are interpreted in the context of the theoretical framework described above. Section~\ref{section_summary} summarizes the principal results of the paper. 

Tabulated results are provided in the HEPData record for this analysis~\cite{hepdata}.

\section{The CMS detector}\label{section_CMS_detector}

The central feature of the CMS apparatus is a superconducting solenoid of 6\unit{m} internal diameter providing a magnetic field of $3.8\unit{T}$. Within the solenoid volume are a silicon pixel and strip tracker, a lead tungstate crystal electromagnetic calorimeter (ECAL), and a brass and scintillator hadron calorimeter (HCAL), each composed of a barrel and two endcap sections. Forward calorimeters extend the pseudorapidity ($\eta$) coverage provided by the barrel and endcap detectors. Muons are measured in gas-ionization detectors embedded in the steel flux-return yoke outside the solenoid. Events of interest are selected using a two-tiered trigger system. The first level (L1), composed of custom hardware processors, uses information from the calorimeters and muon detectors to select events at a rate of around 100\unit{kHz} within a fixed latency of 4\mus~\cite{CMS:2020cmk}. The second level, known as the high-level trigger (HLT), consists of a farm of processors running a version of the full event reconstruction software optimized for fast processing, and reduces the event rate to around 1\unit{kHz} before data storage~\cite{CMS:2016ngn}. This analysis utilized a special data stream collected at HLT rates up to a few \unit{kHz}, described in more detail in Section~\ref{section_bparking}. A more detailed description of the CMS detector, together with a definition of the coordinate system used and the relevant kinematic variables, can be found in Ref.~\cite{CMS_detector_paper}.

\section{The B-parking data sample}\label{section_bparking}

The search uses the CMS B-parking data sample~\cite{EXO-23-007,RKpaper}, which was collected in 2018 using a dedicated data stream to acquire of order $10^{10}$ $\bbbar$ events, corresponding to an integrated luminosity of 41.6\fbinv~\cite{CMS-PAS-LUM-18-002}. A set of single-muon triggers designed to record events containing \PB meson semileptonic decays was used. A variety of requirements were placed on the minimum muon \pt and the muon transverse impact parameter significance \dxysig, where \dxy is the transverse impact parameter of the muon with respect to the beam axis and $\sigma_{\dxy}$ is its uncertainty. More stringent requirements were in place during higher instantaneous luminosity periods and were progressively relaxed at lower instantaneous luminosities to exploit the spare capacity of the CMS data acquisition system and, therefore, maximize the number of recorded \bbbar~events. To maintain good efficiency for muons produced in \PB decays, the muon \pt thresholds were kept low, from 7 to 12\GeV. To control trigger rates, the minimum values used for \dxysig were relatively high, ranging from 3 to 6, thus reducing contamination from prompt muons or muons from charm decays with minimal efficiency penalty for muons from relatively long-lived b hadron decays. Trigger muons were also required to have $\abs{\eta}<1.5$. The purity of this sample, \ie the fraction of events containing a $\bbbar$ pair, is approximately 80\%~\cite{EXO-23-007}. Events in this sample, satisfying the HLT, were collected at a peak rate exceeding 5\unit{kHz}, a much higher trigger rate than is normally possible. The sample was stored temporarily in local buffers and then transferred to permanent storage ("parked"), to be processed at the end of the 2018 data-taking period, when computing resources previously allocated otherwise became available.

\section{Simulated event samples}\label{section_simulation}

Simulated signal event samples are used to design the event selection criteria, to extract the signal shape in the $m(\ell^{\pm}\PGp^{\mp})$ distribution used for fitting the data, and to obtain the signal efficiencies as functions of \mN and \ctau. Signal events involve quantum chromodynamics (QCD) production of a \PQb quark-antiquark pair, followed by the fragmentation of these quarks into a variety of \PB mesons. In the simulation, one of these \PB mesons undergoes either a semileptonic or a leptonic decay, as shown in Fig.~\ref{figure_feynman_diagram}, either of which yields an HNL characterized by its mass, \mN, and its proper mean lifetime, $\tau_{\PN}$. 

Two sets of signal samples are generated: the first corresponds to $\PN$ production in the decays of \Bu, \Bd, and \Bs mesons, while the second corresponds to $\PN$ production in \Bc meson decays. A dedicated simulation of \Bc decays is required because of its low fragmentation fraction compared to other b~hadron species~\cite{PDGReview,Aaij_2019}. For the \main samples, the pp collisions and hadronization are simulated with \PYTHIA~8.230~\cite{Sjostrand:2014zea}, using the underlying event tune CP5~\cite{CMS:2019csb}, the parton distribution functions NNPDF3.1~\cite{NNPDF:2017mvq}, and the SoftQCD:nonDiffractive process (aimed to inclusively reproduce minimum-bias interactions), while the \PB meson and \PN decays are simulated using \EVTGEN~1.30~\cite{LANGE2001152}, with a phase space decay model (PHSP). For the \Bc samples, the pp collisions are first generated with the \textsc{bcvegpy}~2.2b generator~\cite{bcvegpy}, with one \Bc meson per event. The events are then passed to \PYTHIA and \EVTGEN for the hadronization and signal decay, respectively, in a fashion similar to that used for the \main samples. In both cases, the generated events are then propagated through the simulated CMS detector using the \GEANTfour package~\cite{AGOSTINELLI2003250} and eventually reconstructed using the standard CMS software. The presence of extra pp collisions within the same or neighbouring bunch crossings (pileup) is taken into account in these event samples by adding simulated minimum bias events. An event weight is applied to correct the underlying pileup distribution to match that observed in the B-parking data, which has an average of about 25 simultaneous collisions per event. In the signal samples, the same triggers used to record the data are simulated. However, the dependence of the trigger thresholds on the instantaneous luminosity present in data, as described in Section~\ref{section_bparking}, is not accounted for in the simulated samples. Therefore, simulated events are weighted to reproduce this effect, using corrections measured in data with the tag-and-probe method~\cite{CMS_muon} and the benchmark $\PJgy\to\PGm\PGm$ process.
Finally, weights are applied to improve the modelling of the lepton identification algorithm.

Simulated signal event samples are generated with a range of different values of \mN and \ctau to span the accessible parameter space in these quantities. By default, the HNL is assumed to be a Majorana fermion, allowing both $\PN \to \ell^+\PGp^-$ and $\PN\to \ell^-\PGp^+$ decays with equal probabilities. As will be discussed in Section~\ref{section_results}, these simulated samples can also be used to model the Dirac-like case. Signal samples with $1.0\leq \mN\leq 3.0$\GeV are simulated with a step size of 20\MeV for masses $1.0 \leq \mN < 1.5$\GeV, 30\MeV for masses $1.5 \leq \mN < 2.0\GeV$, and 50\MeV for masses $2.0 \leq \mN < 3.0\GeV$. This set contains more than 60 signal mass hypotheses. The step size increases with the tested mass to match the detector resolution. For each value of \mN, one to four samples with different \ctau hypotheses are generated, with $1 < \ctau < 1000\mm$. The grid of the values of \ctau was chosen to be relatively coarse, because, for each \mN, it is possible to obtain samples of events corresponding to different \ctau values by applying an event weight correcting the decay length distribution of the generated sample to a new \ctau value. This procedure, referred to hereafter as \ctau-reweighting, is an important tool in the analysis, as it allows a fine grid of \ctau points to be tested.

Simulated background samples are not required, either for the design of the analysis or the signal extraction.

\section{Event reconstruction}\label{section_reconstruction}

The event reconstruction identifies candidates for the lepton originating from the \PB decay ($\ell_\PB$), the displaced lepton ($\ell^{\pm}$), and the displaced charged pion ($\PGp^{\mp}$). We discuss the criteria used to define each of these reconstructed objects, as well as the procedure used to reconstruct the DV associated with the $\ell^{\pm}\PGp^{\mp}$ system. At this stage, kinematic and topological selections are applied to ensure the presence of a candidate event in the acceptance. Backgrounds are further suppressed using the multivariate discriminator discussed in Section~\ref{section_selection}.

Muons are reconstructed from the tracks measured in the silicon tracker system, together with track segments measured in the muon detectors. The candidates must not have significant associated energy deposits in the calorimeters~\cite{CMS_muon}. Muons that satisfy the HLT requirements for the B-parking data stream are required to have $\pt> 7\GeV$ and $\abs{\eta}<1.5$, while other muons are required to have $\pt>2\GeV$ and $\abs{\eta}<2$. 

The muon candidates originating from the \PB vertex are required to satisfy the following criteria: the associated track in the silicon tracker must correspond to a high-quality track~\cite{CMS_PF,CMS:2010vmp} (at least 5 hits in the silicon tracker and at least 1 hit in the pixel detector), and this track must be matched to at least one muon segment in the muon spectrometer. The track's longitudinal impact parameter, \dz, measured with respect to the primary vertex (PV), defined as the vertex with the largest sum of square of the \pt of its tracks, must be less than 20\unit{cm}, and its transverse impact parameter, \dxy, must be less than 0.3\unit{cm} to be consistent with a typical b~hadron decay.
For displaced muon candidates, a particle-flow reconstruction algorithm~\cite{CMS_PF} is used, and the candidates must satisfy loose displacement criteria: $\dz>15\mum$, $\dxy>10\mum$ and $\dzsig>1.0$, $\dxysig>1.5$.
 
Electrons are reconstructed from energy clusters deposited in the ECAL and the track in the silicon tracker, using a procedure~\cite{CMS_electron} that recovers the energy lost via bremsstrahlung. Electron candidates are required to have $\pt > 1.5\GeV$ and to be within the ECAL barrel, \ie$\abs{\eta}<1.48$. To improve the selection efficiency of real electrons relative to the background, the electrons, either originating from the \PB or \PN vertex, are identified with a multivariate discriminant~\cite{CMS_electron}, based on a boosted decision tree (BDT) architecture. The algorithm uses the properties of the reconstructed electron, including the electromagnetic shower shape, the track-cluster matching, the ECAL/HCAL energy ratio, the track quality, and the track impact parameters with respect to the PV. This BDT discriminant, widely used in CMS, was optimized for electrons with transverse momentum greater than 10\GeV; thus it has been retrained to focus on electrons with comparatively lower \pt, down to 1.5\GeV, and improve its performance in the context of this analysis~\cite{RKpaper}. The displaced electrons are further required to satisfy minimal displacement criteria and to have $\dz>15\mum$ and $\dxy>10\mum$, as well as $\dzsig>1.0$ and $\dxysig>1.5$.  

Displaced pions are reconstructed from high-quality tracks to which the pion mass is assigned. They must have $\pt>1.0\GeV$, $\abs{\eta}<2$, as well as $\dz>50\mum$, $\dxy>50\mum$, and $\dzsig>1.5$, $\dxysig>5.0$. 

Given that the HNL is electrically neutral, the charges of the displaced lepton and pion are required to be opposite. A common DV associated with the intersection of the $\ell^{\pm}$ and $\PGp^{\mp}$ tracks is computed using a kinematic vertex fit~\cite{Prokofiev:2005zz}, which is based on a least mean square minimization with Lagrange multipliers and Kalman filter techniques. The momenta of the displaced lepton and pion are then recomputed using the DV as a constraint, improving the resolution. The $p$-value associated with the fit's $\chi^2$ is required to be greater than 0.01. 

Since only HNLs decaying into visible particles are considered as signal, the reconstructed transverse momentum \ptvec of the $\ell^{\pm}\PGp^{\mp}$ system is aligned with the projection of the direction of flight of the HNL in the transverse plane $\vec{u}_\PN$. However, the exact HNL direction of flight is not known, as we do not identify the \PB decay vertex. Nevertheless, given that, in the most relevant part of the tested phase space, the parent \PB meson travels a significantly shorter distance than an HNL, we assume, with good approximation, that $\vec{u}_\PN$ is represented by the vector joining the luminous region and the DV. The angle $\theta$ between these two vectors $\vec{u}_\PN$ and \ptvec, referred to as the back-pointing angle, peaks at small values and is required to satisfy $\cos\theta > 0.999$. Finally, the significance of the transverse displacement of the DV with respect to the luminous region, \Lxysig, is required to be greater than 15.

As the displacement \Lxy increases, the tracking and DV reconstruction efficiencies decrease, limiting the acceptance for long-lived candidates. The efficiency of the signal candidate reconstruction reaches a few percent for $\Lxy>50\unit{cm}$.

The event reconstruction requirements described above are referred to hereafter as the baseline event selection. This selection consists only of requirements designed to ensure that the two leptons and the charged pion are well reconstructed, the relative charges of the $\ell^{\pm}\PGp^{\mp}$ system are compatible with the signal hypothesis, and the tracks associated with the $\ell^{\pm}\PGp^{\mp}$ system are consistent with a common, displaced decay vertex. For simulated signal events, the reconstructed candidates are matched to the generator-level particles, to ensure that the signal candidates are genuine. For moderate decay lengths, background events passing the baseline selection arise primarily from QCD processes, which can lead to cascade decays of \PB mesons that mimic certain features of the signal signature. In addition, a variety of other combinatorial backgrounds can enter the signal region, and detector reconstruction effects, including particle misidentification, secondary interactions in the tracker material, and pileup, can also play a role. 

\begin{figure}[tbp!]
    \centering
    \includegraphics[width=0.49\textwidth]{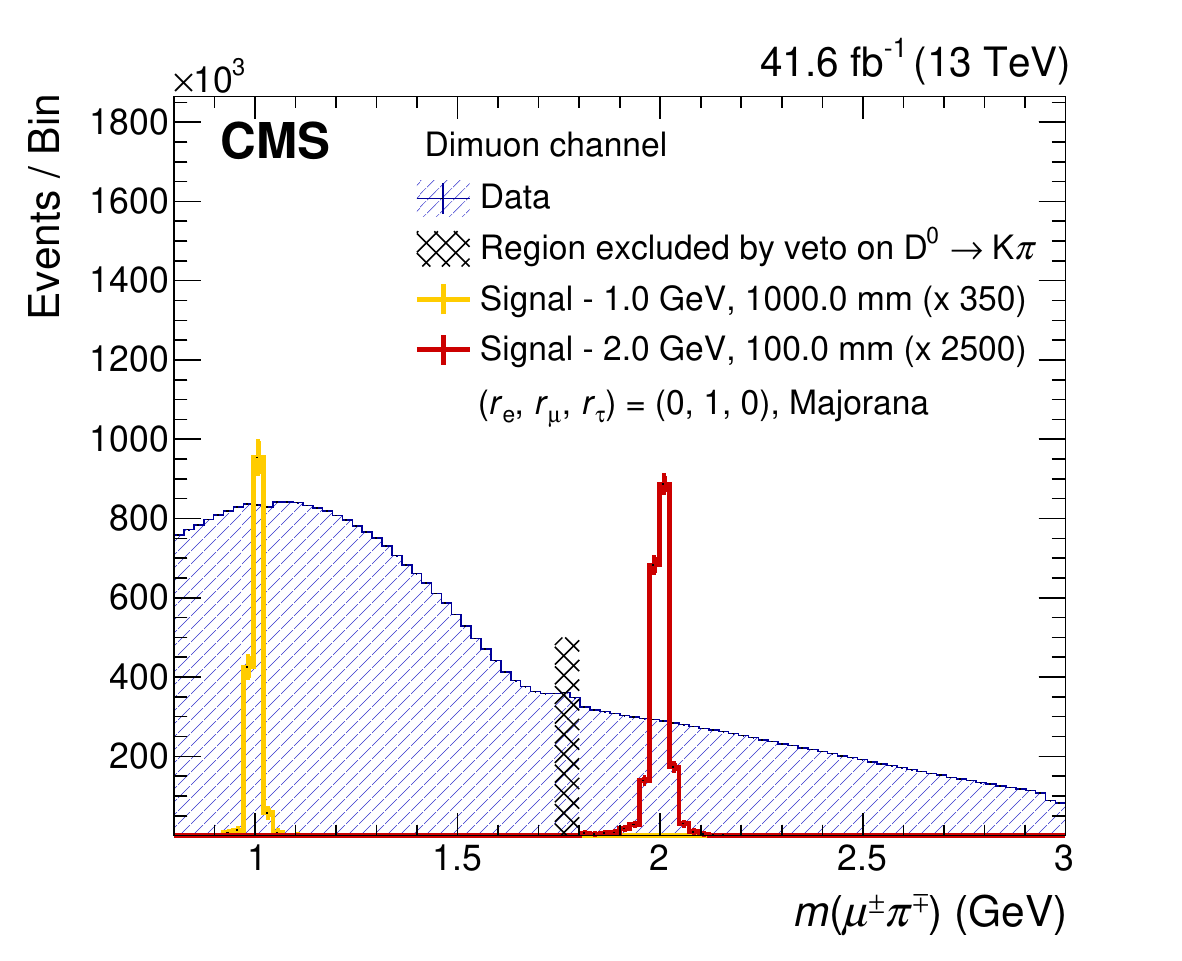}
    \includegraphics[width=0.49\textwidth]{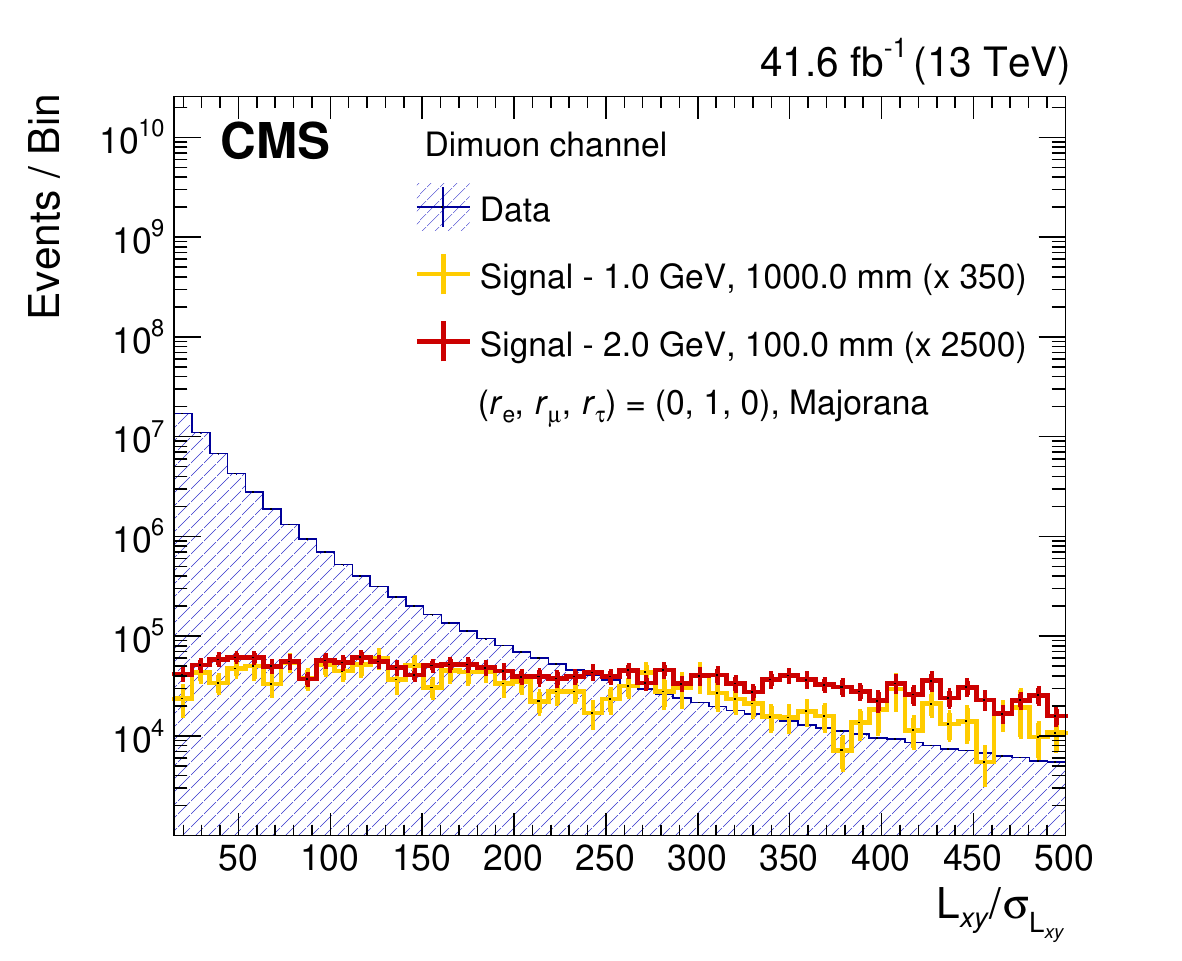}
    \caption{Distribution of the displaced $\PGm^{\pm}\PGp^{\mp}$ invariant mass (left) and \Lxysig (right) in data and in simulated event samples corresponding to two different signal hypotheses, in the Majorana scenario, and with the \PN mixing exclusively with the muon sector: $\mN = 1\GeV$, $\ctau = 1000\mm$, $\V= \Vu = 5.4\times 10^{-4}$; and $\mN = 2\GeV$, $\ctau = 100\mm$, $\V=\Vu=1.7\times 10^{-4}$. The signal distributions are scaled with factors given in the legend. The vertical lines show the statistical uncertainty in each bin. \label{figure_reconstruction}}
    \end{figure}

Figure~\ref{figure_reconstruction} shows distributions of the key variables $m(\ell^{\pm}\PGp^{\mp})$ and \Lxysig for events in the dimuon channel in data and in simulated signal samples, after imposing the baseline event selection. Two examples of the corresponding distributions for signal events in simulation are shown. These signals peak sharply at the assumed masses of the \PN, in this case, $\mN = 1.0$ and 2.0\GeV, as shown in Fig.~\ref{figure_reconstruction} (left). The distributions of \Lxysig are shown in Fig.~\ref{figure_reconstruction} (right). The data distribution falls with increasing values of \Lxysig, as expected from its dominant composition of promptly decaying and short-lived particles.

A number of peaks are observed in the reconstructed invariant mass distributions that are associated with known decay modes of SM particles. The peaks can occur in the two-body systems $\PGm_\PB\PGm^{\pm}$, $\PGm_\PB\PGp^{\mp}$, and $\Pe_\PB \PGp^{\mp}$, which sometimes result from particle misidentification. To suppress backgrounds in which such resonances are present, a set of vetoes is applied, as listed in Table~\ref{table_sm_resonances_veto}. The only residual peaking background in the $\ell^{\pm}\PGp^{\mp}$ mass spectra consists of the $\PDz\to\PK^\pm\PGp^\mp$ process, where one hadron (typically the kaon) is misidentified as a lepton. Therefore, to avoid the possible contamination from two-body $\PDz$ meson decays, no signal extraction, described in detail in Section~\ref{section_signal_extraction}, is performed between 1.74 and $1.80\GeV$. The vetoed region does not centre on the nominal \PDz meson mass owing to mass shift effects related to the final state misidentification.

\begin{table}[bp!]
\centering
\topcaption{List of considered SM resonances and the corresponding vetoes in the various two-particle invariant mass spectra. 
The first seven lines consider any possible opposite-sign pair comprising the lepton originating from the \PB decay and either of the displaced $\ell^{\pm}$ and $\PGp^{\mp}$. Events that fail the veto conditions are removed from the analysis. The last two lines pertain to the displaced $\ell^{\pm}\PGp^{\mp}$ candidate and indicate that the signal extraction is not performed and exclusion limits are not provided for $m_\PN$ in the vetoed regions. The presence of misidentified particles is also indicated. For the \PDz meson vetoes, the mass range is adjusted to account for the incorrect mass hypothesis assigned to the misidentified particle.\label{table_sm_resonances_veto}}
\begin{tabular}{ccccc}
\hline
Mass spectrum                & Process                    & Veto (GeV)                               & Categories & Misidentification \\
\hline
$m(\PGm_\PB\PGm^\pm)$        & $\Pgf\to\PGm\PGm$  & $\abs{m(\PGm_\PB\PGm^\pm) - 1.02} > 0.01$ & OS         & 0                 \\
                             & $\PJgy\to\PGm\PGm$ & $\abs{m(\PGm_\PB\PGm^\pm) - 3.10} > 0.15$ & OS         & 0                 \\
                             & $\Pgy\to\PGm\PGm$  & $\abs{m(\PGm_\PB\PGm^\pm) - 3.69} > 0.08$ & OS         & 0                 \\ [\cmsTabSkip]
  
$m(\PGm_\PB\PGp^\mp)$        & $\PJgy\to\PGm\PGm$ & $\abs{m(\PGm_\PB\PGp^\mp) - 3.10} > 0.05$ & SS         & 1 misid. \PGp     \\
                             & $\PDz\to\PK\PGp$   & $\abs{m(\PGm_\PB\PGp^\mp) - 1.76} > 0.05$ & SS         & 1 misid. \PGm     \\ [\cmsTabSkip]

$m(\Pe_\PB\PGp^\mp)$         & $\PJgy\to\Pe\Pe$   & $\abs{m(\Pe_\PB\PGp^\mp) - 3.10} > 0.05$  & SS         & 1 misid. \PGp     \\
                             & $\PDz\to\PK\PGp$   & $\abs{m(\Pe_\PB\PGp^\mp) - 1.76} > 0.05$  & SS         & 1 misid. \Pe      \\ [\cmsTabSkip]
$m(\PGm^\pm\PGp^\mp)$        & $\PDz\to\PK\PGp$   & $\abs{m(\PGm^\pm\PGp^\mp) - 1.77} > 0.03$ & all        & 1 misid. \PGm     \\ [\cmsTabSkip]
$m(\Pe^\pm\PGp^\mp)$         & $\PDz\to\PK\PGp$   & $\abs{m(\Pe^\pm\PGp^\mp) - 1.77} > 0.03$  & all        & 1 misid. \Pe      \\ 
\hline
\end{tabular}
\end{table}

\section{Event categories and selection}\label{section_selection}
After the baseline selection described in Section~\ref{section_reconstruction}, events are assigned to categories according to (i) three ranges in the significance of the transverse decay length, \Lxysig, of the $\ell^{\pm}\PGp^{\mp}$ system, (ii) two possibilities for the relative sign of the charged leptons, (iii) two ranges in the invariant mass of the $\ell_\PB\ell^\pm\PGp^\mp$ system, and (iv) the dimuon and mixed-flavour channels introduced in Section~\ref{section_introduction}. Table~\ref{table_categorization} lists the categories associated with these quantities, as well as the three lepton-flavour categories allowed for the selected $\ell_\PB\ell$ combinations. Combining these criteria leads to an event classification in terms of 24 mutually exclusive categories. The category based on \Lxysig provides a way to increase the sensitivity to different \ctau hypotheses. Because the dominant backgrounds arise from promptly or quasi-promptly decaying QCD processes (relative to the \PN lifetime), the background falls with \Lxysig, as seen in Fig.~\ref{figure_reconstruction} (right). The relative sign of the candidate leptons originating from the \PB and \PN decay vertex is also a useful quantity for event classification: signal events can contain either SS or OS leptons, while background events are predominantly OS, \eg \PB to \PD cascade decays. Finally, the categorization in the \bmass provides a way to discriminate signals originating from a \Bu, \Bd, or \Bs meson from those originating from a \Bc meson. In fact, heavy neutrinos from \Bc mesons are produced almost exclusively through leptonic decays, where all three final state particles, $\ell_\PB$, $\ell^{\pm}$, and $\PGp^{\mp}$, are reconstructed, allowing the reconstruction of the \Bc invariant mass.

\begin{table}[tbh!]
\centering
\topcaption{Summary of the event categorization. The events are classified into 24 mutually exclusive categories. \label{table_categorization}}
\begin{tabular}{ccc}
\hline
Quantity & Label & Definition \\
\hline
\Lxysig & low \Lxysig & $\Lxysig < 50$ \\
& medium \Lxysig & $50 < \Lxysig < 150$ \\
& high \Lxysig & $\Lxysig > 150$ \\ [\cmsTabSkip]
Relative lepton sign & OS & $\ell_\PB$ charge $\neq$ $\ell$ charge \\
& SS & $\ell_\PB$ charge = $\ell$ charge \\ [\cmsTabSkip]
\bmass & low \bmass & $\bmass < 5.7\GeV$ \\
& high \bmass & $\bmass > 5.7\GeV$ \\ [\cmsTabSkip]
Flavour channel & dimuon & $\ell_\PB\ell = \PGm\PGm$ \\
& mixed-flavour & $\ell_\PB\ell = (\PGm\Pe,\, \Pe\PGm)$ \\
\hline
\end{tabular}
\end{table}

To further improve the sensitivity of the search, a multivariate discriminator is constructed using a wide range of event variables that distinguish statistically between signal and background processes. Because the signal distributions of these variables generally depend on \mN, a method is needed to allow the discriminator quantity to incorporate a functional dependence on \mN. This capability is provided by a parametric neural network (pNN)~\cite{parametricNN}. Such a neural network (NN) has, in addition to a set of input variables, $\vec{x}$, a set of discrete parameters, $\vec{\theta}$, and it  returns a score that is a function of the full variable set $f\left(\vec{x},\ \vec{\theta}\right)$. For this analysis, only a single parameter, \mN, is used. By incorporating \mN as a parameter, the need for separate trainings for different values of \mN (more than 60 in this analysis) is avoided, and the pNN behaves smoothly for mass values between those used in the training sample. The mass parameters chosen for the training are $\mN = 1.0$, 1.5, 2.0, and 3.0\GeV for both signal and background. The signal signatures also depend on the value of \ctau. However, this variable is not used as an additional parameter of the pNN, as it was found that training the pNN separately in the different categories of \Lxysig ensures good performance for all of the \ctau hypotheses.

The pNN is trained separately in each of the categories defined in Table~\ref{table_categorization}. The baseline event selection described in Section~\ref{section_reconstruction} is applied to all training samples. The simulated signal event samples for $\mN = 1.0$, 1.5, 2.0, and 3.0\GeV are aggregated into a common signal training sample in which the discrete $\mN$ values are provided to the pNN as a parameter. Because simulating the large cross section background processes is too resource intensive, the background training sample is taken from data but comprises less than one thousandth of all the available data events. The use of a limited data set ensures that the contamination from a hypothetical \PN signal would be negligible, while not impacting the performance of the pNN in any significant way. To properly train the pNN, it is important to ensure that, for each value of the parameter $\mN$, the events in the background sample have kinematic properties that correspond to the background events in the \invmass invariant mass region near $\mN$. Therefore, selected background events lie within a $\pm10\sigma$ window around the $\mN$ mass peak, where $\sigma$ is the mass resolution obtained in the reconstruction of the simulated signal events. 

The pNN is trained using input variables that provide a good discrimination between signal and background. These variables, which are discussed in more detail below, are
\begin{enumerate}
\item Transverse momenta: $\pt(\ell_\PB)$, $\pt(\ell^{\pm})$, $\pt(\PGp^{\mp})$.
\item Invariant-masses: $m(\ell_\PB\PGp^{\mp})$, $m(\ell_\PB\ell^{\pm})$, $m(\ell_\PB\ell^{\pm}\PGp^{\mp})$.
\item Track separation in the $\eta$-$\varphi$ space (where $\varphi$ is the azimuthal angle), $\Delta R \equiv \sqrt{\smash[b]{(\Delta\eta)^2+(\Delta\varphi)^2}}$: $\Delta R(\ell_\PB, \ell^{\pm})$, $\Delta R(\ell_\PB, \PGp^{\mp})$.
\item Displaced vertex properties: $\cos\theta$, fit $p$-value.
\item Displacement-related quantities: $\Lxy/{\sigma_{\Lxy}}$ and \dxysig of the pion.
\item Track-related information: number of layers of the CMS silicon pixel and strip tracker traversed by the lepton(s) and pion from the DV. 
\item Lepton isolation, defined in a cone of $\Delta R$ smaller than 0.3 around the lepton momentum vector~\cite{CMS_muon, CMS_electron}.
\end{enumerate}

\begin{figure}[tbp!]
\centering
\includegraphics[width=0.49\textwidth]{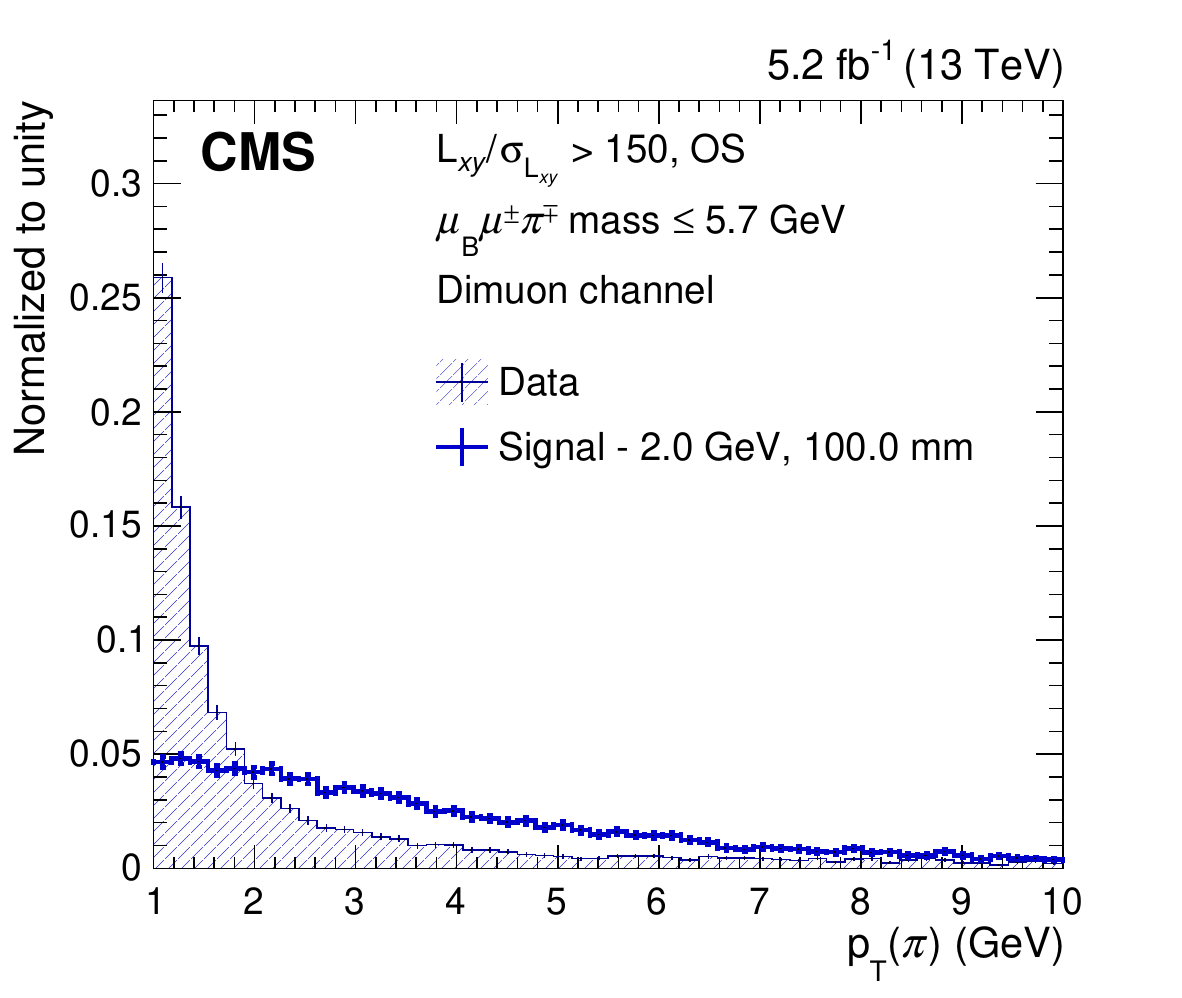}
\includegraphics[width=0.49\textwidth]{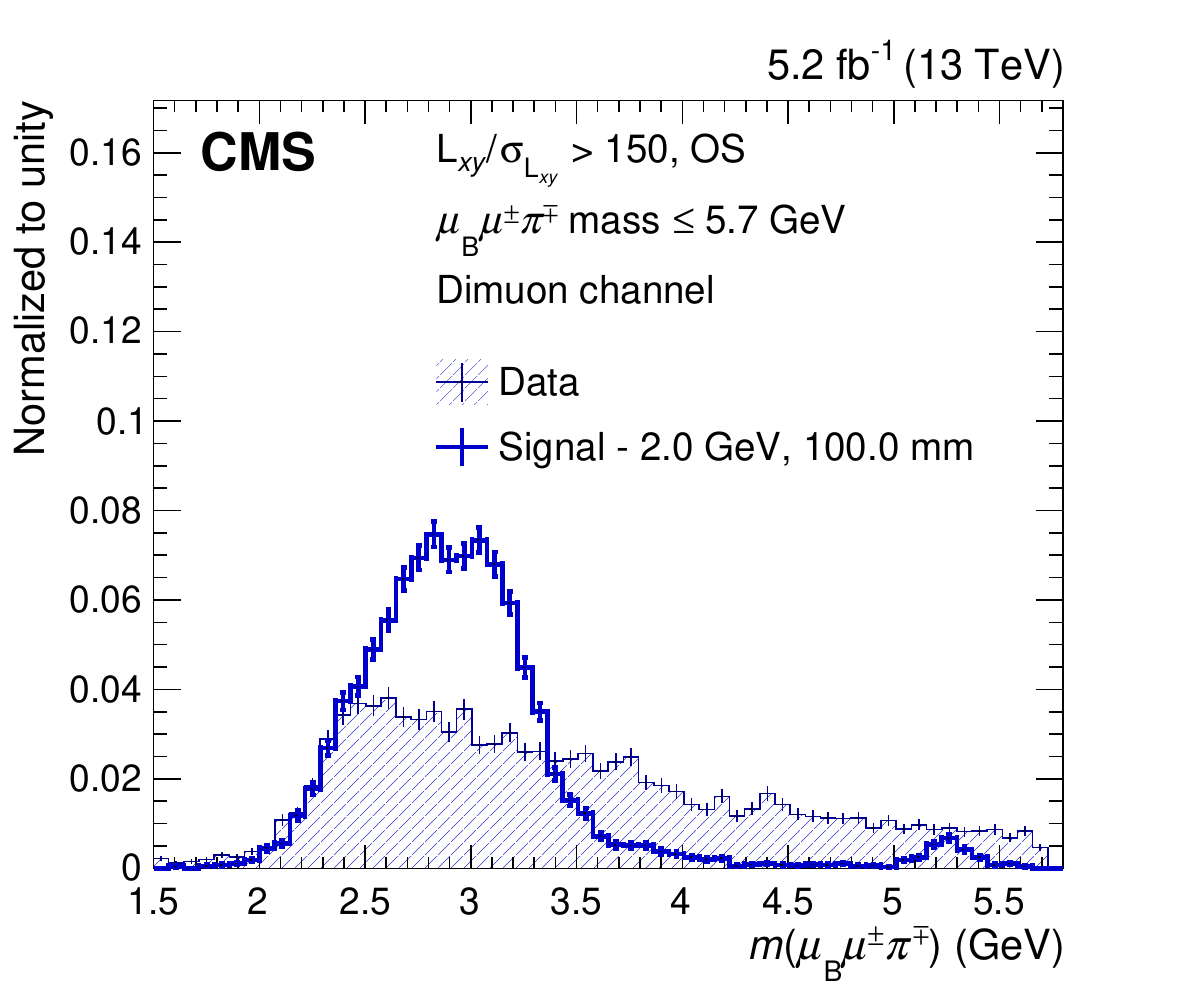}
\includegraphics[width=0.49\textwidth]{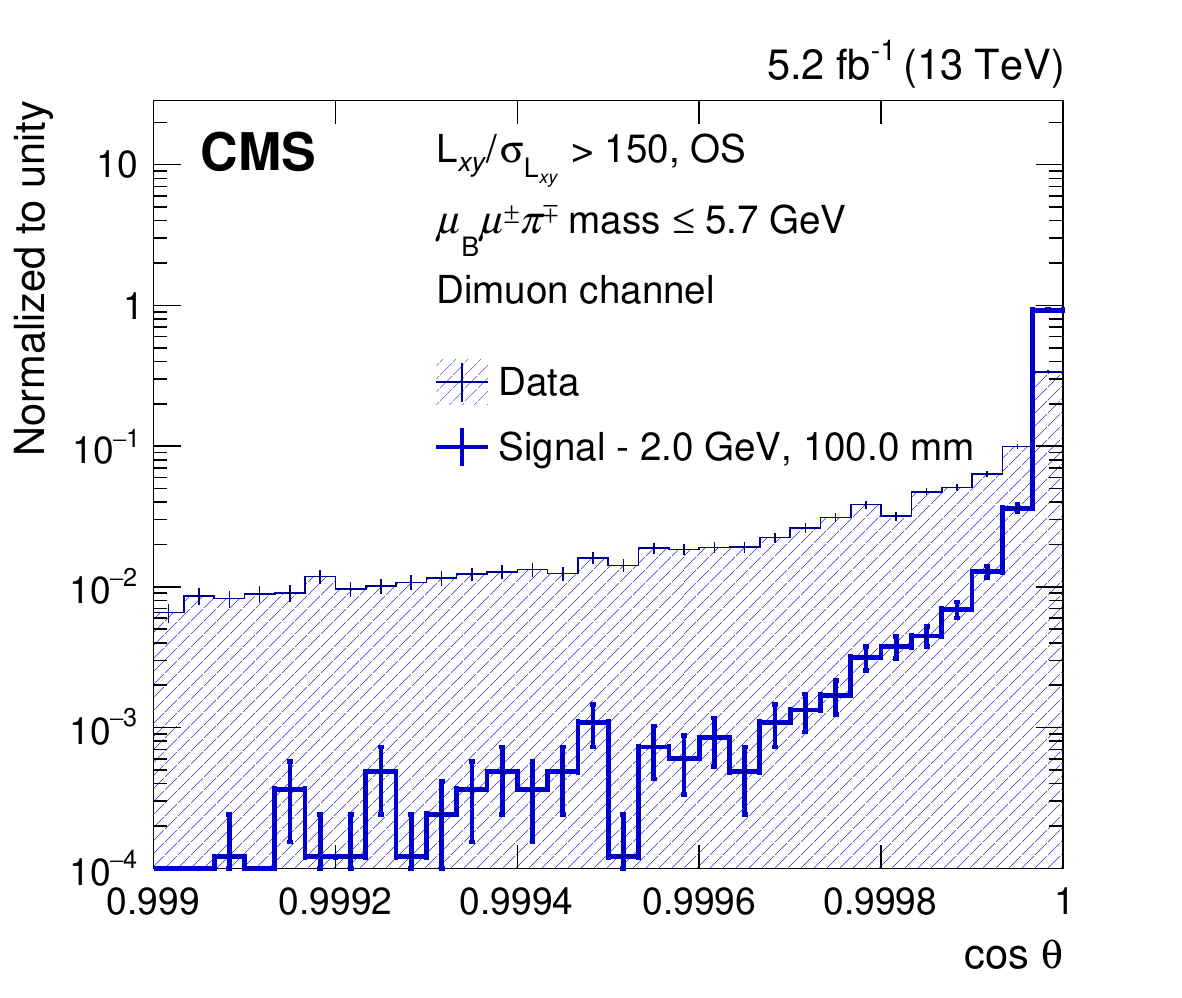}
\includegraphics[width=0.49\textwidth]{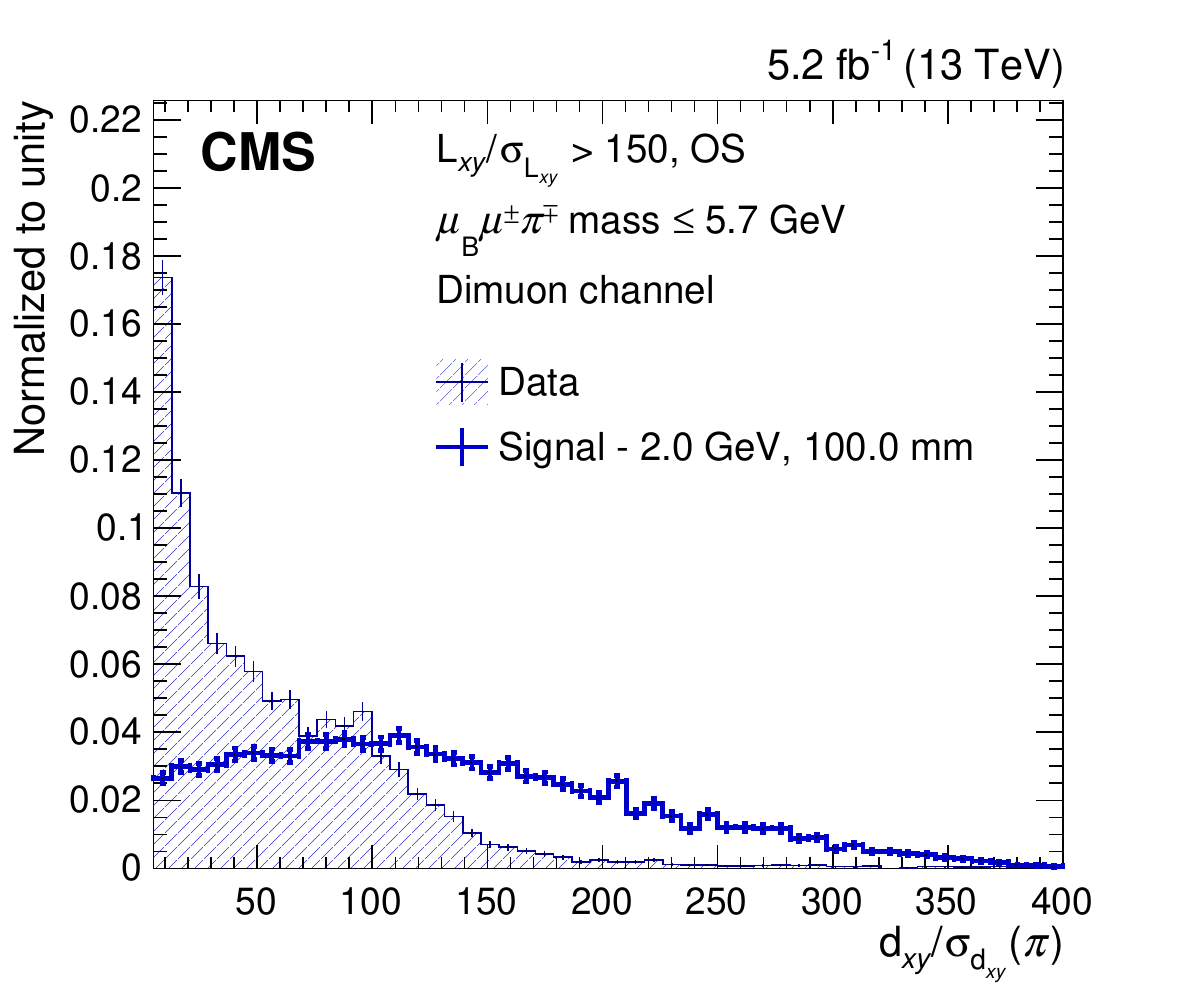}
\caption{The distributions of the pion \pt (upper left), $m(\PGm_\PB\PGm^\pm\PGp^\mp)$ (upper right), $\cos\theta$ (lower left), and pion \dxysig (lower right) are shown for data, as well as for a signal hypothesis of $\mN=2\GeV$ and $\ctau=100\mm$. The data correspond to an integrated luminosity of 5.2\fbinv and are selected in the mass window of size $10\sigma$ around $\mN=2\GeV$. The distributions, which are normalized to unit area, are shown for the dimuon channel in category with high \Lxysig, OS, and low \bmass. The vertical lines show the statistical uncertainty in each bin.
\label{figure_pNN_features}}
\end{figure}

Comparisons of signal and background distributions for some of these variables are given in Fig.~\ref{figure_pNN_features}, for representative signals (i) in the dimuon channel, with $\mN=2\GeV$ and $\ctau=100\unit{mm}$ and (ii) in the category with high \Lxysig, OS, and low \bmass. Because the \pt distribution of the background is typically softer than that of the signal, as can be seen in Fig.~\ref{figure_pNN_features} (upper left) for the displaced pion, the transverse momentum of the three particles in the final state is added to the variable set. The separation power offered by these quantities depends on the value of \mN. For example, the \pt of the displaced particles $\ell^\pm$ and $\PGp^\mp$ gets harder as \mN gets larger, offering even greater discrimination between the signal and the background.

The set of discriminating variables listed above includes the invariant masses $m(\ell_\PB\ell^\pm)$, $m(\ell_\PB\PGp^\mp)$, and $m(\ell_\PB\ell^\pm\PGp^\mp)$. Figure~\ref{figure_pNN_features} (upper right) shows the distribution of $m(\ell_\PB\ell^\pm\PGp^\mp)$. For lower values of \mN, the \PB meson decays predominantly via semileptonic processes, while for large values, the semileptonic decay rate falls rapidly and the leptonic rate becomes dominant. In the case where \PB decays via a semileptonic process, $m(\ell_\PB\ell^\pm\PGp^\mp)$ does not peak at the mass of the \PB meson, $m_\PB$, as the accompanying hadronic system \X, introduced in Section~\ref{section_introduction}, is not reconstructed. In the leptonic case, a peak at $m_\PB$ is observed, which provides further background rejection.

Because the momenta of the HNL's decay products typically have similar directions in the laboratory frame, the pNN is also trained on the track separation in $\eta$-$\phi$ space, $\Delta R(\ell_\PB, \ell^{\pm})$, and $\Delta R(\ell_\PB, \PGp^{\mp})$. The quantity $\Delta R(\ell^{\pm}, \PGp^\mp)$ is not added to the training variable set, as, otherwise, the signal mass \mN could be learned by the pNN during training because of the addition of the transverse momenta.

The pNN uses two quantities that characterize the \PN decay vertex, providing additional sensitivity to the \ctau hypothesis within each bin of \Lxysig. These quantities are $\cos\theta$, where $\theta$ is the back-pointing angle defined in Section~\ref{section_reconstruction}, and the DV fit $p$-value. Figure~\ref{figure_pNN_features} (lower left) shows signal and background distributions of the quantity $\cos\theta$, which typically peaks at 1 more steeply for signal than for background. As shown in the list above, the pNN is also trained on the \Lxysig and the \dxysig of the pion. The distribution of the latter quantity is shown in Fig.~\ref{figure_pNN_features} (lower right), in which we can see that the long-lived signal reaches larger values compared to the background.
 
Because the signal is a long-lived particle, additional information on the tracks is provided to the pNN. In particular, the number of layers in the silicon pixel and strip tracker traversed by the charged particles provides good signal-to-background discrimination. Larger values of \ctau typically correspond to fewer tracker layers traversed by the daughter lepton and charged pion from the \PN decay. Finally, the lepton isolation is added to the training variable set, as leptons in a signal event tend to be more isolated than leptons in a background event. This quantity is defined, for both the muon and the electron, as the energy sum of the charged and neutral particles lying in a cone within a distance $\Delta R=0.3$ from the lepton, from which the contribution from pileup particles is subtracted.

For each mass hypothesis, \mN, the pNN score assigned to an event can be interpreted as an estimate of the probability that the event was produced by the signal process with HNL mass \mN. This quantity is therefore used to enhance the signal significance with respect to the background. In each mass window, introduced in Section~\ref{section_signal_extraction}, and in each category, listed in Table~\ref{table_categorization}, the events are selected by requiring the pNN score to be greater than 0.99. This threshold was obtained from a dedicated study in which the performance of the tested thresholds was evaluated based on the median expected exclusion limit of benchmark signal hypotheses. To avoid bias, this optimization was performed without viewing the complete data sample. 

The pNN architecture is that of a fully connected NN with three layers:
\begin{enumerate}
\item Input layer: contains as many nodes as the number of input features, plus one extra node for the mass parameter \mN.
\item Hidden layer: contains 64 nodes and is activated by the rectified linear unit (ReLU) function~\cite{4082265, keras}, defined as ReLU($x$) = max(0, $x$), with $x\in\mathbb{R}$.
\item Output layer: returns the pNN score, normalized to unity through the sigmoid activation function. 
\end{enumerate}

The training is performed using the \textsc{TensorFlow} package~\cite{tensorflow}, with the \textsc{Keras} interface~\cite{keras} and the \textsc{Adam} optimizer~\cite{adam}. The training aims at minimising the cross-entropy loss function, and the weights are updated using batch gradient descent. The batch size is chosen to be 32 and the initial learning rate to be 0.01. The input features are normalized using the \textsc{RobustScaler}~\cite{scikit-learn} before the training.

\begin{figure}[tbh!]
\centering
\includegraphics[width=0.49\textwidth]{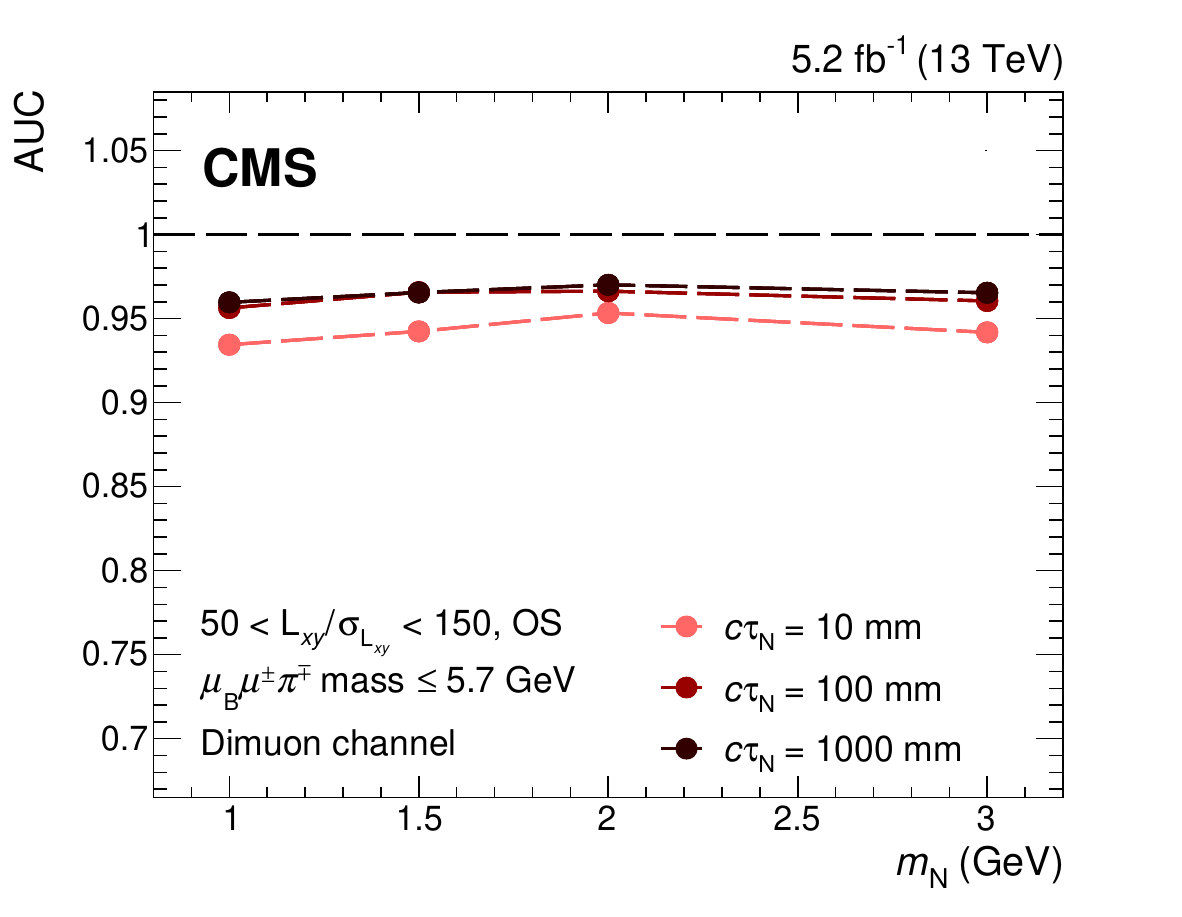}
\includegraphics[width=0.49\textwidth]{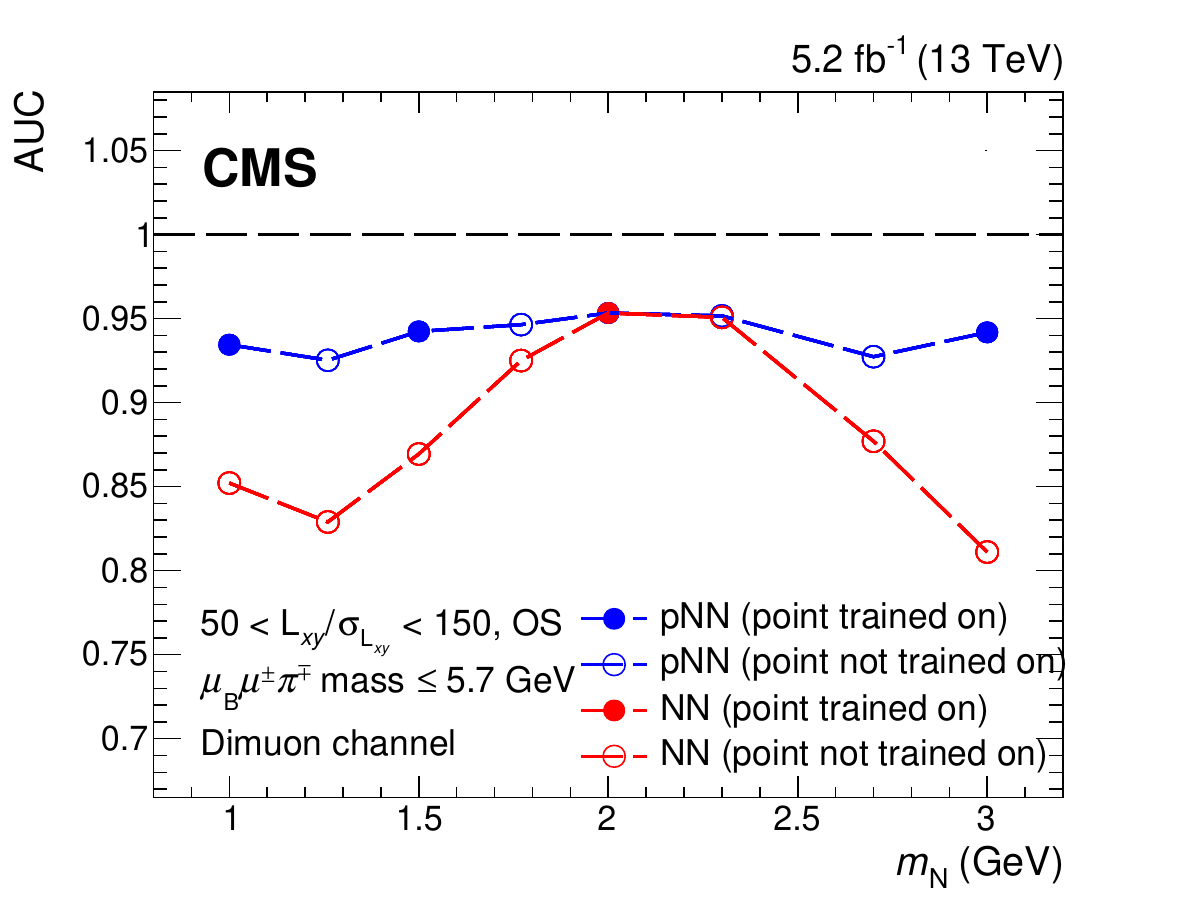}
\caption{(Left)  Performance of the pNN as a function of signal mass for events in the $50<\Lxysig<150$, OS, and low \bmass category in the dimuon channel. The performance is shown by the AUC curve, where a value at unity corresponds to a perfect separation between signal and background. The different coloured curves correspond to different \ctau hypotheses. (Right) Validation of the use of a pNN for intermediate \mN mass points, for events in the $50<\Lxysig<150$, OS and low \bmass category in the dimuon channel: a pNN trained on masses $\mN = 1.0$, 1.5, 2.0, and 3.0\GeV (blue) and a NN trained on mass $\mN=2\GeV$ (red). All the points have $\ctau=10\mm$. The full circles correspond to mass points on which the pNN and NN were trained on, while the open circles show mass points that have not been trained on. 
\label{figure_pNN}}
\end{figure}

The performance of the pNN is quantified using the area under the ROC curve (AUC), as shown, for example, in Fig.~\ref{figure_pNN}~(left) for events in the dimuon channel, and in the category with medium \Lxysig, OS, and low \bmass. The AUC is given as a function of the signal mass for different \ctau hypotheses. The large values of the AUC show that the pNN performs well for the various signal hypotheses. The pNN performs well for a wide range of \ctau values, with marginal degradation for signals with a smaller decay length.

To validate the use of pNNs instead of regular NNs trained for specific masses, we compare the AUC values obtained for the two approaches. The results of this study are summarized in Fig.~\ref{figure_pNN}~(right) for events in the dimuon channel, in the medium \Lxysig, OS, and low \bmass category. The first set of AUC values shown corresponds to the pNN described above and trained on signal masses of $\mN = 1.0$, 1.5, 2.0, and 3.0\GeV. The second set of AUC values was obtained using a NN with the same architecture and training conditions as the pNN, but instead trained only on the  mass $\mN=2\GeV$ sample. The different points have $\ctau=10\mm$ and are obtained from a corresponding simulated sample, without the use of the \ctau-reweighting procedure introduced in Section~\ref{section_simulation}. We conclude that the pNN and NN yield the same AUC score for $\mN=2\GeV$, the common mass they are trained on. Furthermore, the pNN performance remains approximately constant for masses lying between those used for the training, in contrast to the performance loss of the NN for masses other than $\mN=2\GeV$.

\section{Signal normalization}\label{section_normalization}

The number of expected signal events, $N_{\text{sig}}$, in the flavour channel $\ell_\PB\ell = (\PGm\PGm,\, \Pe\PGm,\, \PGm\Pe)$, for a given \PN hypothesis specified by \mN and \ctau, and mixing scenario specified by $\vec{r} = (\fe, \fu, \ft)$, is given by
\begin{equation}
N_{\text{sig}}\left(\ell_\PB\ell, \mN, \ctau, \vec{r} \right) = \frac{\sigma_{\PBpm}^\mathrm{eff}}{f_u} \mathcal{L} \sum_{{\PQq}} F_\PN^{\PQq} \left(\ell_\PB\ell, \mN, \ctau, \vec{r} \right)    \epsilon_{\text{sig}}^{\PQq}\left(\ell_\PB\ell, \mN, \ctau \right), 
\label{equation_signal_yield} 
\end{equation}
where $\sigma_{\PBpm}^\mathrm{eff}$ is the effective total cross section for both $\PB^+$ and $\PB^-$ meson production, extrapolated from the measured fiducial cross section of the process \control in the phase space covered by the trigger conditions described in Section~\ref{section_bparking}; $f_u$ is the fragmentation fraction of the charged \PB meson \Bu; $\mathcal{L}$ is the integrated luminosity for the B-parking data sample; the quark flavour index, \PQq, runs over (\PQu, \PQd, \PQs, \PQc); $F_\PN^\PQq$ is the product of the \PN production rate and decay branching fractions associated with a signal originating from the decay of a meson \Bq; and $\epsilon_{\text{sig}}^{\PQq}$ is the associated signal efficiency, incorporating the acceptance, reconstruction, trigger, and selection efficiencies. The signal factor, $F_\PN^\PQq$, in Eq.~(\ref{equation_signal_yield}) is defined as 
\begin{equation}\label{equation_FN}
\hspace{-0.1cm} F_\PN^{\PQq} \left(\ell_\PB\ell, \mN, \ctau, \vec{r} \right)=  
    \underbrace{\sum_{\mathrm{~}\X_{\PQb}}  f_\PQq \frac{\widetilde{\Gamma}(\PB_\PQq \to \ell_\PB \PN \X_{\PQb})}{\Gamma(\PB_\PQq)} r_{\ell_\PB} \V}_{\PN\,\text{production}}       
    \underbrace{\frac{\widetilde{\Gamma}\left(\PN \to \ell \PGp\right)}{\fe\widetilde{\Gamma}_\Pe\left(\PN\right) + \fu\widetilde{\Gamma}_\PGm\left(\PN\right) + \ft\widetilde{\Gamma}_\PGt\left(\PN\right)} r_{\ell}\vphantom{\sum_{{\PQq},\mathrm{~}\X_{\PQb}}  f_{\PQq}    \frac{\widetilde{\Gamma}(\PB_{\PQq} \to \ell \PN \X_{\PQb})}{\Gamma(\PB_{\PQq})} r_{\ell} \V}}_{\PN\,\text{decay}}.
\end{equation}
In the \PN production term, the sum runs over the possible hadronic systems, $\X_{\PQb}$, associated with the decay of the meson \Bq; $f_{\PQq}$ is the fragmentation fraction of the meson $\PB_{\PQq}$; $\Gamma(\PB_{\PQq})$ is the SM $\PB_{\PQq}$ decay width~\cite{PDGReview}; and $\widetilde{\Gamma}(\PB_{\PQq} \to \ell_\PB \PN \X_{\PQb})$ is the partial decay rate for inclusive production of the \PN with lepton of flavour $\ell_\PB$ (as computed from Ref.~\cite{bondarenko}) divided by $\abs{V_{\ell_\PB\PN}}^2$, which is factored out and written explicitly as $r_{\ell_\PB}\V$. In the \PN decay term, the factor $\widetilde{\Gamma}(\PN \to \ell \PGp)$ is the partial decay width with \Vl factored out, and $\widetilde{\Gamma}_\alpha(\PN)$ is the total decay width for an \PN decaying into a charged or neutral lepton of flavour $\alpha=(\Pe, \PGm, \PGt)$, with \Valpha factored out. In the ratio between \Vl and \Valpha, the factor \V cancels out, leaving the mixing ratio $r_{\ell}$, in the numerator, and the mixing ratios, $\fa$, in the denominator. The \PN decay widths are computed based on Ref.~\cite{bondarenko}. 

The quantities $\V$, \ctau, and $\vec{r}$ are linked to each other by the following relation:
\begin{equation}\label{eq:tau_vs_coupling}
\frac{1}{\ctau} = \V\left(\fe\widetilde{\Gamma}_\Pe\left(\PN\right) + \fu\widetilde{\Gamma}_\PGm\left(\PN\right) + \ft\widetilde{\Gamma}_\PGt\left(\PN\right)\right),
\end{equation}
which implies that different combinations of $\vec{r}$ and $\V$ yield the same \PN lifetime \ctau. This property underpins the interpretations of the results in various flavour mixing, as well as in the Majorana or Dirac-like scenarios, discussed in Section~\ref{section_results}.

\begin{figure}[tbp!]
\centering
\includegraphics[width=0.6\textwidth]{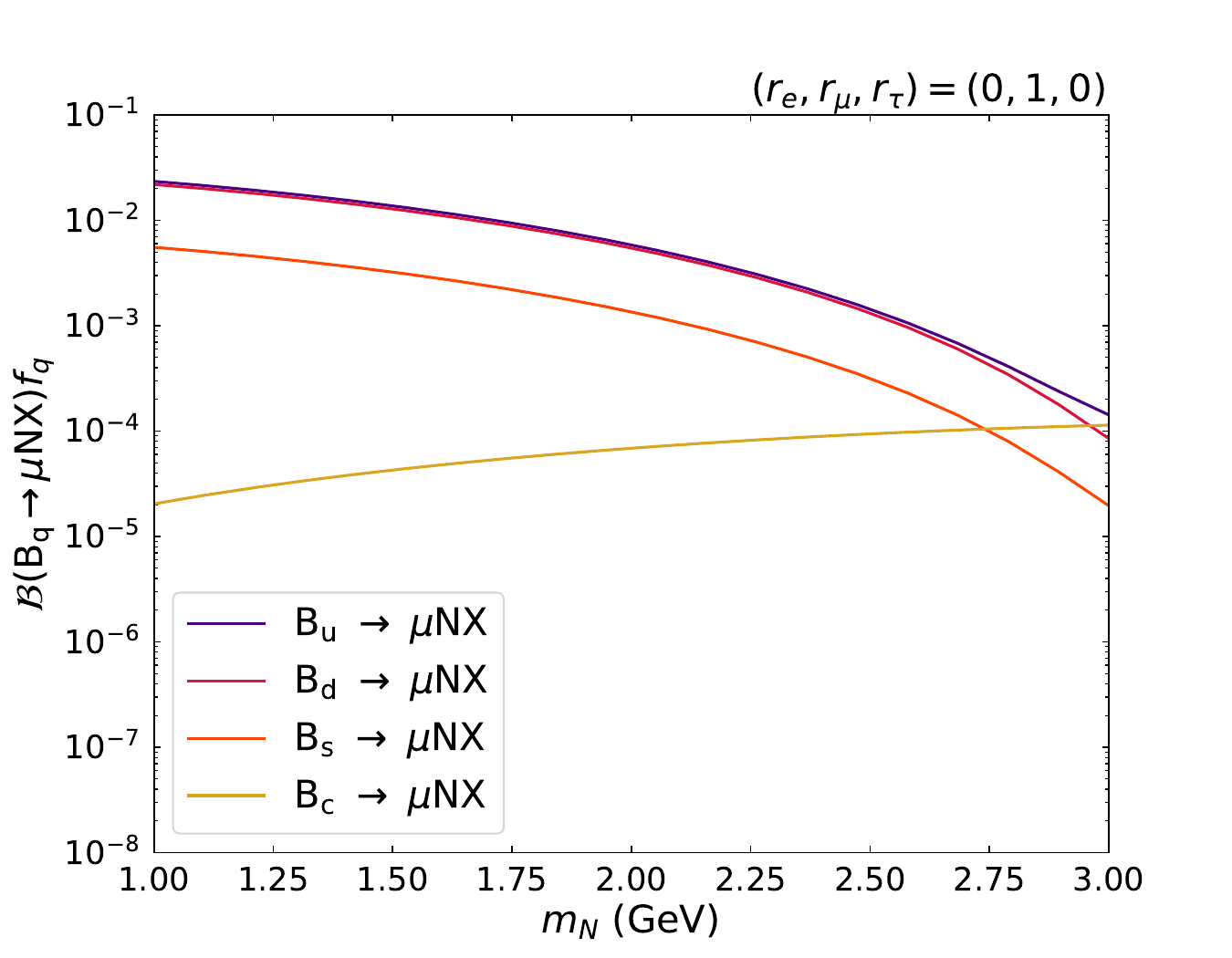}
\caption{Branching fractions as functions of $\mN$ for $\PB_\PQq\to\PGm\PN\X$ decays, $\PQq = (\PQu, \PQd, \PQs, \PQc)$, multiplied by the corresponding fragmentation fraction, $f_{\PQq}$. Both leptonic and semileptonic decays are considered. The results are shown for the mixing scenario $\V=\Vu=1$. The branching fractions are computed based on the method described in Ref.~\cite{bondarenko}. \label{figure_production_BR}}
\end{figure}

The decay branching fractions multiplied by the fragmentation fractions, $f_{\PQq}$, for the four $\PB$ meson species are shown in Fig.~\ref{figure_production_BR} as a function of \mN for the scenario in which $\PN$ can only couple to muons. For masses approaching $\mN=3\GeV$, the \Bc meson contribution becomes significant, despite its low fragmentation fraction, because of the predominance of leptonic decays and the larger \Bc mass. For smaller masses, the dominant contribution is from the \Bu, \Bd, and \Bs mesons. The branching fractions for the process $\PN\to\PGm^\pm\PGp^\mp$ range from 22\% down to 2.4\% for \mN masses from 1.0 to 3.0\GeV, respectively, in the scenario where the \PN couples exclusively to muons.
  
\begin{figure}[tbp!]
\centering
\includegraphics[width=0.6\textwidth]{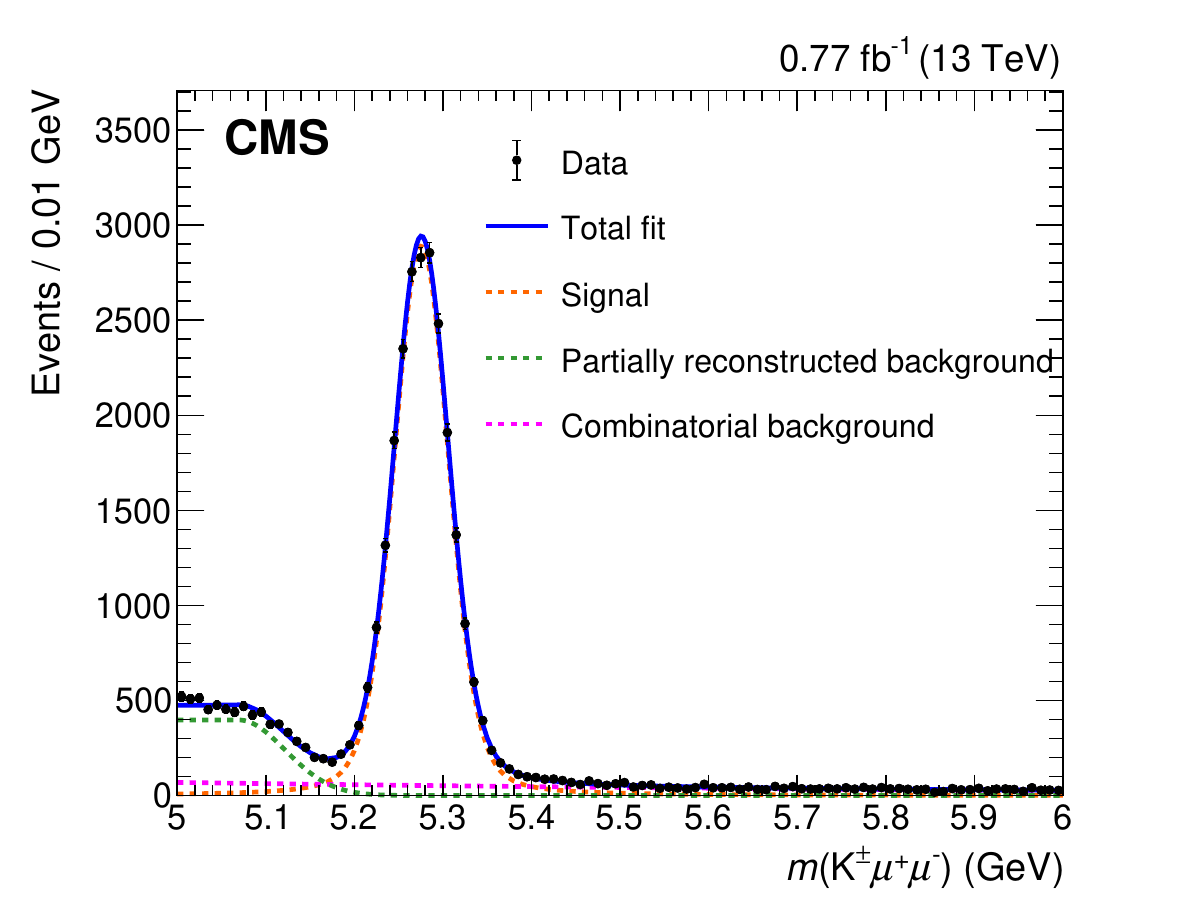}
\caption{Distribution of the $\PK^\pm\PGm^+\PGm^-$ invariant mass for a luminosity of 0.77\fbinv. A large signal is observed at the $\PB_{\PQu}$ mass. The blue curve shows the fit to signal plus background, while the orange, green, and red curves show the contributions from the signal, composite background, and combinatorial background, respectively. \label{figure_fit_control_region}}
\end{figure}

The total effective cross section, $\sigma_{\PBpm}^\mathrm{eff}$, is extrapolated from the measurement of the cross section for the SM control process \control, performed in the fiducial phase space imposed by the trigger requirements of the B-parking sample, discussed in Section~\ref{section_bparking}. In particular, the leading muon must have $\pt>7\GeV$ and $\abs{\eta}<1.5$. We note that a similar measurement has already been performed by CMS, within a defined fiducial phase space~\cite{sigmaBpaper}. In a manner similar to Eq.~(\ref{equation_signal_yield}), the expected number of control events is computed as the product of the cross section, $\sigma_{\PBpm}^\mathrm{eff}$; the relevant SM branching fractions; the integrated luminosity; and the efficiency of the SM decay, accounting for the acceptance, selection, and trigger effects. The efficiency is obtained from a simulation of the control process. The associated sample is simulated using the same generator conditions as for the signal samples described in Section~\ref{section_simulation}, such that the measurement of $\sigma_{\PBpm}^\mathrm{eff}$ using the control process can be extrapolated to the signal process. The cross section $\sigma_{\PBpm}^\mathrm{eff}$ is then obtained by measuring the number of control events in data. The event yield is extracted from the data using an extended unbinned maximum likelihood fit, shown in Fig.~\ref{figure_fit_control_region}. The total fit accounts for three components: (i) the signal peak, (ii) the background arising from misreconstructed b~hadron decays, such as $\PB \to \PJgy + \mathrm{hadrons}$ (partially reconstructed background), and (iii) the background  originating from a spurious combination of a \PJgy meson and an uncorrelated track (combinatorial background). The sample used for this purpose has an integrated luminosity of 0.77\fbinv, a fraction of that of the full event sample. However, the statistical uncertainty in the yield for the control process is much smaller than the systematic uncertainty (discussed in Section~\ref{section_systematics}). No theoretical uncertainty for the \PYTHIA model used for the extrapolation is considered here, since the fiducial phase space selected in this ancillary measurement is close to that covered by the signal in the analysis, such that this uncertainty cancels out in the signal normalization procedure. The measured value is $\sigma_{\PBpm}^\mathrm{eff} = (572.0 \pm 4.9\stat \pm 85.8\syst)$\unit{$\mu$b}. 

\section{Signal extraction}\label{section_signal_extraction}

The signal event yield for each heavy-neutrino mass hypothesis, $\mN$, is extracted from a simultaneous parametric fit to the $m(\ell^{\pm}\PGp^{\mp})$ distributions in data in the different categories listed in Table~\ref{table_categorization}. The baseline selections, the 2-body resonance vetoes listed in Table~\ref{table_sm_resonances_veto}, and the requirement on the pNN score to be greater than 0.99 are all applied before these fits are performed. Because \mN is not predicted and can assume any value allowed by the kinematics of the decays under study in this analysis, we perform the fits to the data in a series of sliding $m(\ell^{\pm}\PGp^{\mp})$ mass windows centred around a set of closely spaced trial mass values. The separation of these trial values is about twice the mass resolution, $\sigma$, and the window around each trial mass value is $\pm 10\sigma$. The values of $\sigma$ range approximately from 9\MeV to 25\MeV for signal masses of 1.0 and 3.0\GeV, respectively. 

The sum of signal plus background models is fitted to the mass distributions. The signal mass shape is parametrized, in all the categories, and for all \PN hypotheses, with a double-sided Crystal Ball function~\cite{doubleCB1,doubleCB2}. While the parameters describing the power-law tails of the function are fixed, the signal mass resolution, $\sigma$, increases linearly with the signal mass $\mN$. 

As anticipated in Section~\ref{section_reconstruction}, the background is generally falling; however, its shape is not known a priori, and it is allowed to vary in the different mass windows and categories. The background shape is studied prior to the final fit in the sidebands, defined as the regions outside the signal window of size $\pm 3\sigma$ around the mass hypothesis \mN. The parameters controlling the background shapes used in the fit are free-floating and are treated as uncorrelated for the 24 categories and the different mass windows.

To account for possible background shape variations, the discrete profiling method~\cite{Dauncey_2015} is used. This method employs a set of alternative background shape functions in each mass window and category. The method also provides a way to incorporate a systematic uncertainty associated with the choice of functional form for the background shape by treating this choice as a discrete nuisance parameter (NP). During the fit, at each point of the scan over the signal strength, the profile-likelihood is evaluated for each function indexed by NP and the best fit one is chosen. The final set of functions considered consists of three families: power-law polynomials, Laurent-series polynomials, and a sum of exponentials. Initially, each family contemplates all functions of any order $N$.

To avoid overfitting, the maximum order, $N_{\rm max}$, of the function that is considered in each family, in the sense that a higher order function would not improve the fit, is assessed using a Fisher test~\cite{FTest}. To determine the value of $N_\mathrm{max}$ for a given family of functions, successive orders are tested iteratively. If the function at order $N$ provides an adequate fit, or $N$ is the lowest order, the function of order $N+1$ is tested by computing $\Delta\mathrm{NLL} = 2(\mathrm{NLL}_N - \mathrm{NLL}_{N+1})$, with NLL$_{N}$ being the negative log-likelihood of the fit with the function of order $N$. According to Wilk's theorem~\cite{10.1214/aoms/1177732360}, $\Delta$NLL behaves asymptotically as a $\chi^2$ with one degree of freedom, allowing one to compute the associated $p$-value of the test as $p^\mathrm{F} = \mathrm{p}(\chi^2>\Delta\mathrm{NLL})$. If $p^\mathrm{F}<0.05$, the function of order $N+1$ is supported by the data, and the same test will be run for the next-order function. If not, then the testing stops and $N_\mathrm{max} = N+1$. For each family, the functional form of order $N_\mathrm{max}$ is directly added to the final set of functions, as well as the functions with lower order, provided that their associated $\chi^2$-goodness of fit probability exceeds 0.01. 

\begin{figure}[tbh!]
\centering
\includegraphics[width=0.6\textwidth]{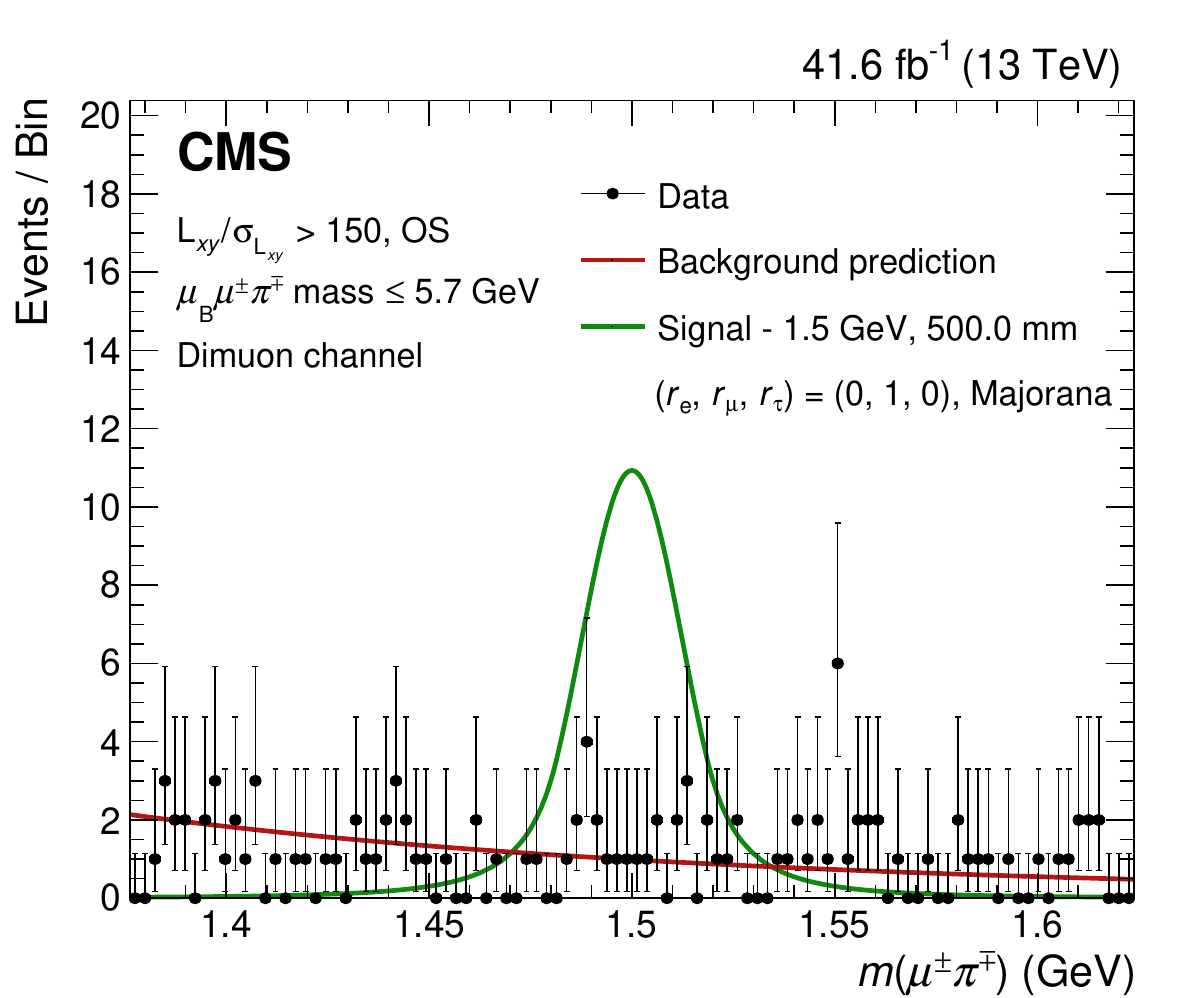}
\caption{Distribution of the $\PGm^\pm\PGp^\mp$ invariant mass in the mass window around 1.5\GeV in the high \Lxysig, OS, and low \bmass category in the dimuon channel. The result of the background-only fit to the data (red) is shown together with the mass distribution expected from a Majorana signal with $\mN=1.5\GeV$ and $\ctau=500\mm$, for the case in which the \PN mixes with the muon sector only (green). \label{figure_prefit}}
\end{figure}

Figure~\ref{figure_prefit} shows an example of a background-only fit in a mass window around 1.5\GeV, in the high \Lxysig, OS, and low \bmass category in the dimuon channel. The functional form of the background is the one in the set returned as having the best $\chi^2$-goodness of fit. The invariant mass distribution expected for a representative signal is overlaid. 

\section{Systematic uncertainties}\label{section_systematics}
 
The systematic uncertainty in the signal event yield due to the modelling of the background is assessed using the discrete profiling method in the signal extraction, as described in Section~\ref{section_signal_extraction}. All of the other uncertainties are summarized in Table~\ref{table_systematics} and are associated with the shape parametrization of the signal invariant mass and the normalization of the signal event yield. These uncertainties are discussed below and are small compared to the uncertainties arising from the size of the event sample or from the fit procedure.

The first source of systematic uncertainty listed in Table~\ref{table_systematics} is associated with parametrization of the shape of the signal invariant mass distribution. As discussed in Section~\ref{section_signal_extraction}, the signal mass distribution is parametrized using a double-sided Crystal Ball function. In each flavour channel, the parameters of the power-law tails are fixed and the resolution is parametrized linearly in the signal mass, while being treated inclusively between the different \ctau hypotheses; the flavour channels; and the categories in \Lxysig, relative lepton sign, and \bmass. A systematic uncertainty of 15\%, obtained by studying the relative change on the median expected limit when varying some of the fit parameters, is applied to account for the possible effects of the choice of signal parametrization. While all the resolution and parameters of the tails are fixed in the final fit, the mean of the distribution has a freedom of 0.1\% and 2\% for the dimuon and mixed-flavour signals, respectively, to account for the lepton energy uncertainties.   

The other systematic uncertainties listed in Table~\ref{table_systematics} are associated with how the signal event yield is related to the theoretical parameters of interest, as obtained in Eq.~(\ref{equation_signal_yield}). The main sources of systematic uncertainty in the determination of the effective \PBpm cross section, $\sigma^{\mathrm{eff}}_{\PBpm}$, (discussed in Section~\ref{section_normalization}) arise from the choice of the model used in the likelihood fits, the reconstruction of the \PB meson decay vertex, and the muon trigger. These contributions amount to a total uncertainty of 15\% in $\sigma^{\mathrm{eff}}_{\PBpm}$. Signal events from \Bc decays (in the high \bmass categories) are subject to an additional uncertainty of 24\% in the $f_\PQc$ fragmentation because of the currently limited experimental accuracy of the $f_\PQc$ measurement~\cite{Aaij_2019}.
The remaining uncertainties listed in Table~\ref{table_systematics} are associated with the signal efficiency, $\epsilon_\text{sig}^\mathrm{\PQq}$. These are

\begin{itemize}
\item Uncertainty in the signal efficiency arising from the pNN selection. To estimate this uncertainty, we consider the control process \control. This selection, described in Section~\ref{section_selection}, is applied in both data and simulation. The residual difference between the pNN output score distribution between data and simulation is taken as the systematic uncertainty. It ranges between 5 and 20\%, depending on the event category and it is assigned equally to the dimuon and the mixed-flavour channels, as the pNN is flavour agnostic.
\item Uncertainty associated with the matching, described in Section~\ref{section_reconstruction}, between a reconstructed signal candidate and the underlying generated signal candidate. This uncertainty is 5\%.
\item Uncertainty in the signal efficiency arising from potential mismodelling of tracking- and vertexing-related quantities that affect measurements of the decay vertices. These are accounted for with a systematic uncertainty of 5\%, based on Ref.~\cite{WHNLpaper}.
\item  Uncertainties in the signal efficiency arising from correction factors applied to the simulated signal event samples. Several types of such scale factors are applied to improve the agreement between the actual and simulated behavior of the detector, including efficiency scale factors for electrons, muons, and the trigger. The uncertainties in the corrections are 5\% and 1\% for the trigger and lepton identification scale factors, respectively. These uncertainties are computed by varying the respective scale factor weights up and down by $1$ standard deviation and studying the impact on the number of expected signal events. Lepton scale factors are computed using a tag-and-probe method~\cite{CMS_muon}.
\end{itemize}

The systematic uncertainties discussed above are modelled as log-normal uncertainties. Apart from the uncertainty in $\sigma_{\PBpm}^\mathrm{eff}$ and \fc, the uncertainties are treated as uncorrelated across the different analysis categories. 

Finally, the statistical uncertainty associated with the limited size of the simulated signal samples and the per-event \ctau-reweighting is taken into account. These uncertainties are assumed to be described by $\Gamma$ distributions.

\begin{table}[tbp!]
\centering
\topcaption{Sources of systematic uncertainty affecting the expected signal event yield. The ranges given correspond to the uncertainties across the different event categories. The uncertainty in the integrated luminosity is not reported as it is incorporated in the uncertainty in the cross section measurement used to normalize the signal. \label{table_systematics}}
\begin{tabular}{lc}
\hline
Source                                & Value (\%)  \\
\hline          
Signal shape                          & 15          \\
$\sigma_{\PBpm}^\mathrm{eff}$         & 15          \\
\fc                                   & 24          \\
Signal selection                      & 5--20       \\
Limited simulated signal sample size  & $<$15       \\
Matching                              & 5           \\
Tracking efficiency                   & 5           \\
Trigger scale factors                 & 5           \\
Muon identification scale factors     & 1           \\
Electron identification scale factors & 3           \\[\cmsTabSkip]
Total                                 & $<$42       \\
\hline
\end{tabular}
\end{table}

\section{Results and interpretation}\label{section_results}

\begin{figure}[tbh!]
\centering
\includegraphics[width=0.49\textwidth]{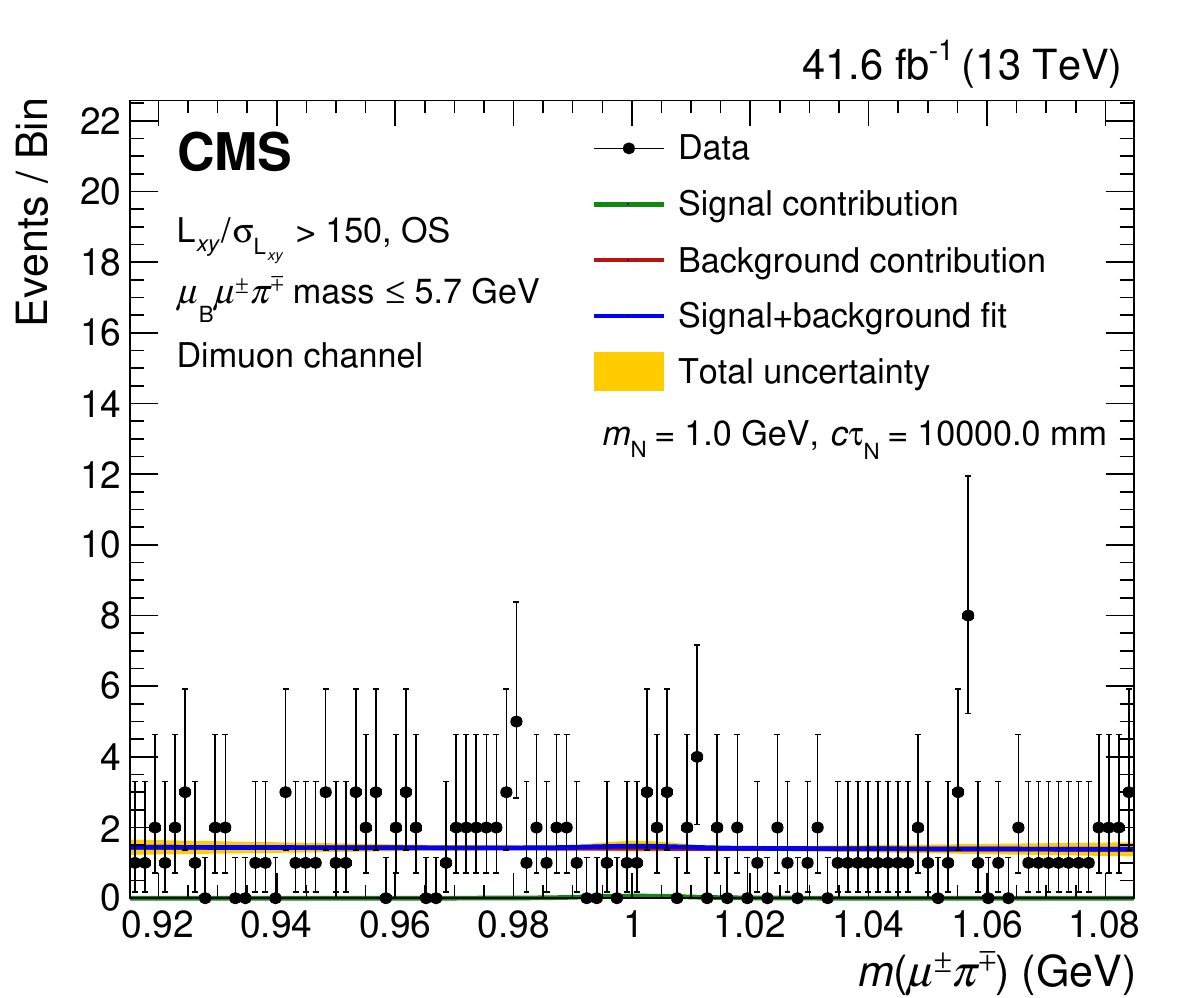}
\includegraphics[width=0.49\textwidth]{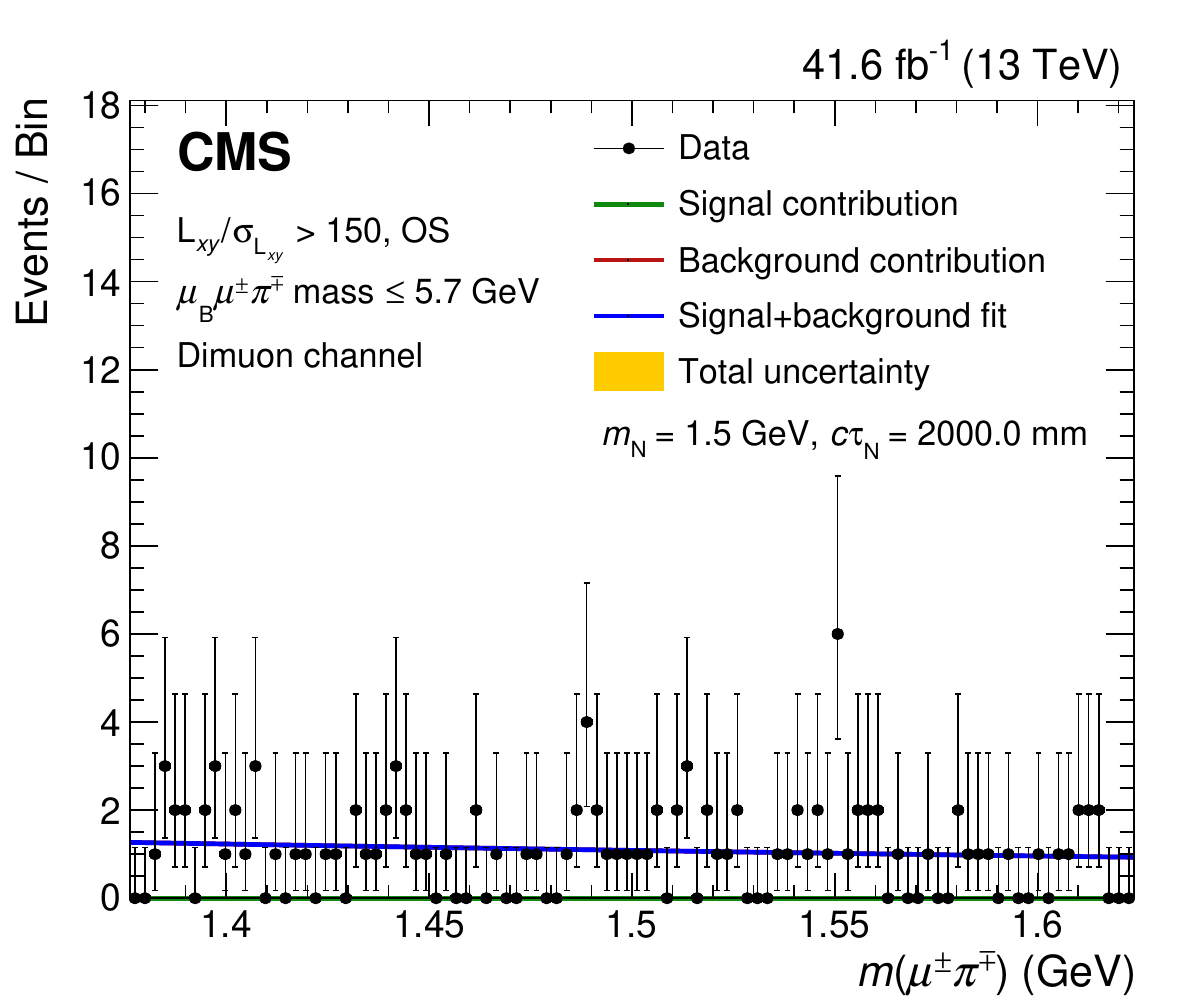}
\includegraphics[width=0.49\textwidth]{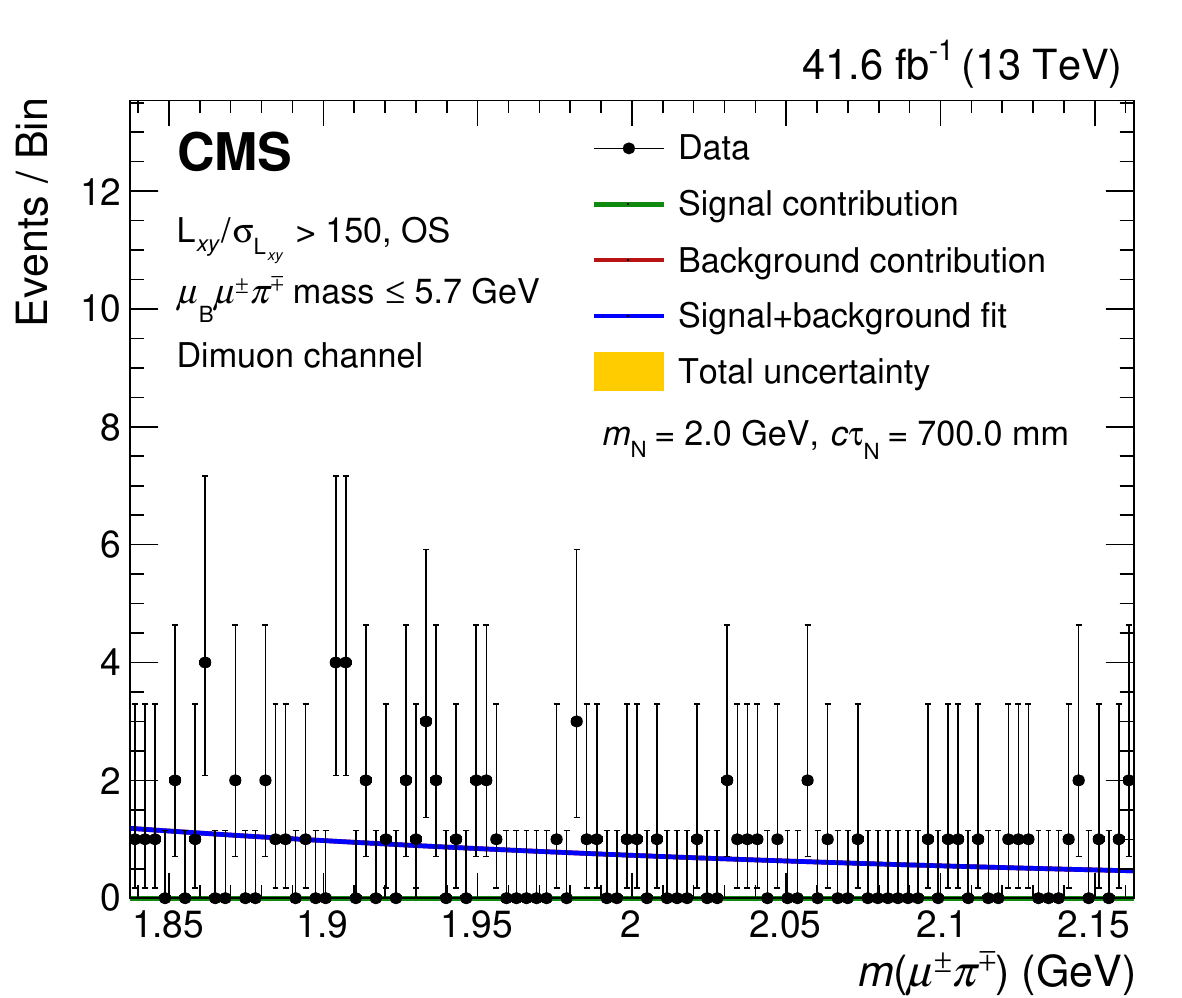}
\includegraphics[width=0.49\textwidth]{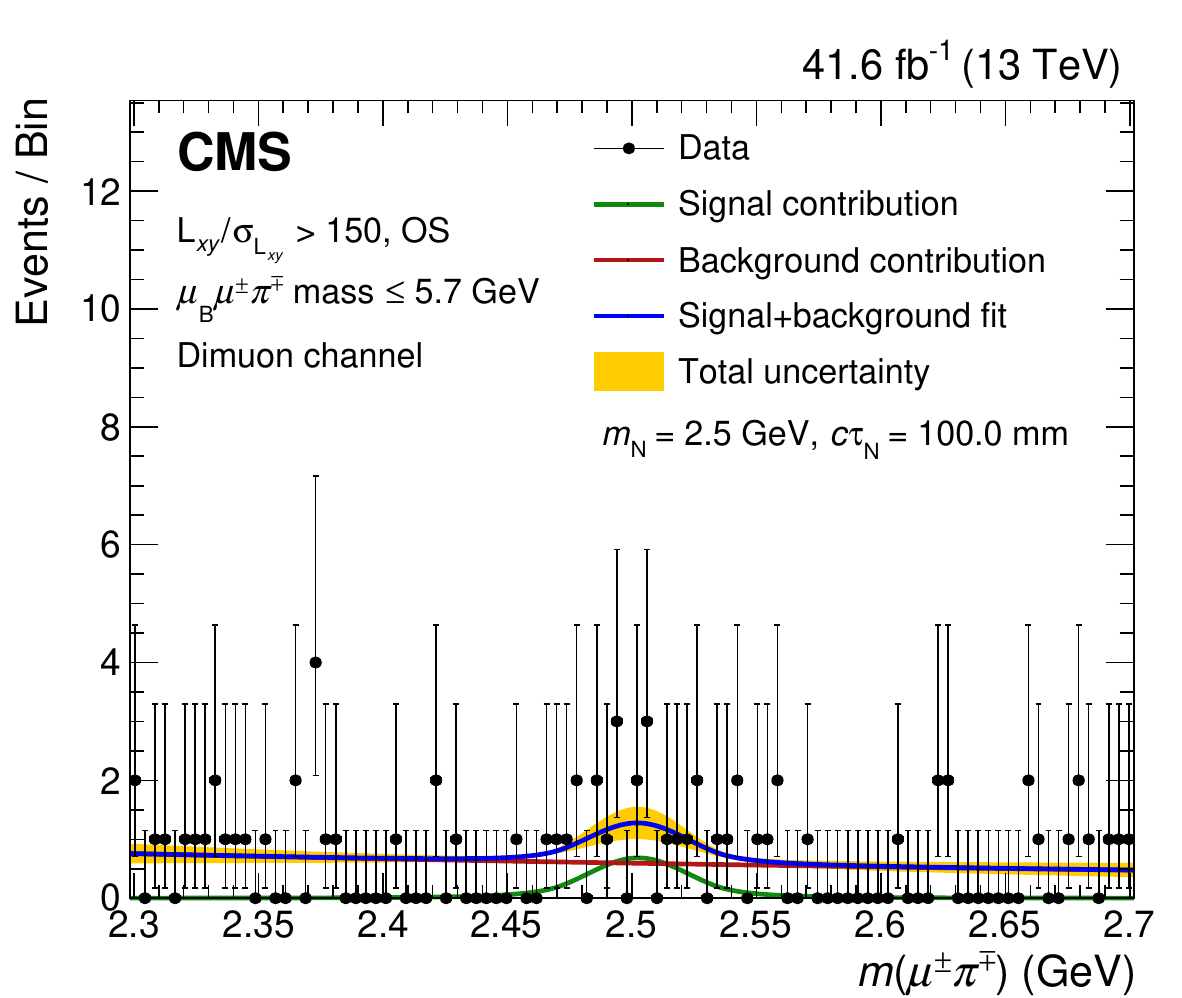}
\caption{Fits of the mass distribution for a signal of mass $\mN=1.0$\GeV (upper left), 1.5\GeV (upper right), 2.0\GeV (lower left), and 2.5\GeV (lower right), in the high \Lxysig, OS, and low \bmass category of the dimuon channel. The blue curve corresponds to signal-plus-background fit, while the green and red curves indicate its individual signal and background components, respectively. The yellow band shows the total post-fit systematic plus statistical uncertainty. \label{figure_postfit}}
\end{figure}

For each mass window around a given signal mass hypothesis, \mN, the signal yield is extracted from the data using the procedure explained in Section~\ref{section_signal_extraction} using the uncertainties summarized in Section~\ref{section_systematics}. No significant excesses over the background-only expectations are observed. As an example, Fig.~\ref{figure_postfit} shows the $\ell^{\pm}\PGp^{\mp}$ invariant mass distributions in the dimuon channel for the high \Lxysig, OS, and low \bmass category; the mass windows shown are centred around $\mN = 1.0$, 1.5, 2.0, and 2.5\GeV. For these four mass windows, the figure shows the corresponding fits to the data assuming signal hypotheses for the cases $\ctau = 10000.0$, 2000.0, 700.0, and 100.0\mm, respectively. These values are close to the exclusion limit when the HNL mixes exclusively with the muon sector, which is the scenario with the greatest sensitivity. The plots in the mass windows around $\mN = 1.0$, 1.5, and 2.0\GeV are examples showing data that do not present a signal-like structure and are representative of most of the distributions considered. The plot in the mass window around $\mN=2.5\GeV$ is an example chosen to show a signal-like fluctuation in the fit region. 

For each value of $\mN$, the results are interpreted using Eqs.~(\ref{equation_signal_yield}) and~(\ref{equation_FN}), which relate the observed number of signal events, $N_{\text{sig}}$, to the quantity \V. Upper limits on the signal yields at the 95\% confidence level (\CL) are used to establish corresponding limits on the quantity \V, assuming a variety of different mixing scenarios specified by the ratios (\fe, \fu, \ft) at a fixed proper mean lifetime $\tau_\PN$. These limits are obtained with the combination of the different event categories listed in Table~\ref{table_categorization}. For each signal hypothesis (\mN, \ctau) and each mixing scenario, an unbinned maximum likelihood function, $\mathcal{L}(r, \vec{\theta})$, is constructed using the signal strength, $r$, as the parameter of interest and the full set of NPs, $\vec{\theta}$, that account for the different sources of uncertainty. 

The limits are derived using the \CLs criterion~\cite{CLscriterion} in the asymptotic approximation~\cite{Cowan2011}. A given signal hypothesis is considered to be excluded if a signal strength equal to (or larger than) unity is not compatible with the observed data at 95\% \CL. A scan of \V on a fine grid is performed using the \ctau-reweighting methodology described in Section~\ref{section_simulation}. 

Limits are derived for both the Majorana and Dirac-like nature of the HNL. As discussed in Section~\ref{section_simulation}, the signal simulation samples are generated for the Majorana case. 
In the Dirac-like scenario, lepton flavour is conserved, and only OS $\ell_\PB^{\pm}\ell^{\mp}$ pairs are allowed in the signal process. In the OS categories, for a given \mN and \ctau point, the expected event yield for the Dirac-like scenario can be obtained from that of the Majorana scenario considering the fact that the decay widths of a Dirac-like \PN are half of those of a Majorana \PN~\cite{bondarenko}, resulting in twice the yields in the former case than in the latter. Finally, another consequence of the change of the decay widths is that, for a given \mN and \ctau hypothesis, the \V in the Dirac-like scenario is twice as large as in the Majorana scenario.

\begin{figure}[tbh!]
\centering
\includegraphics[width=0.49\textwidth]{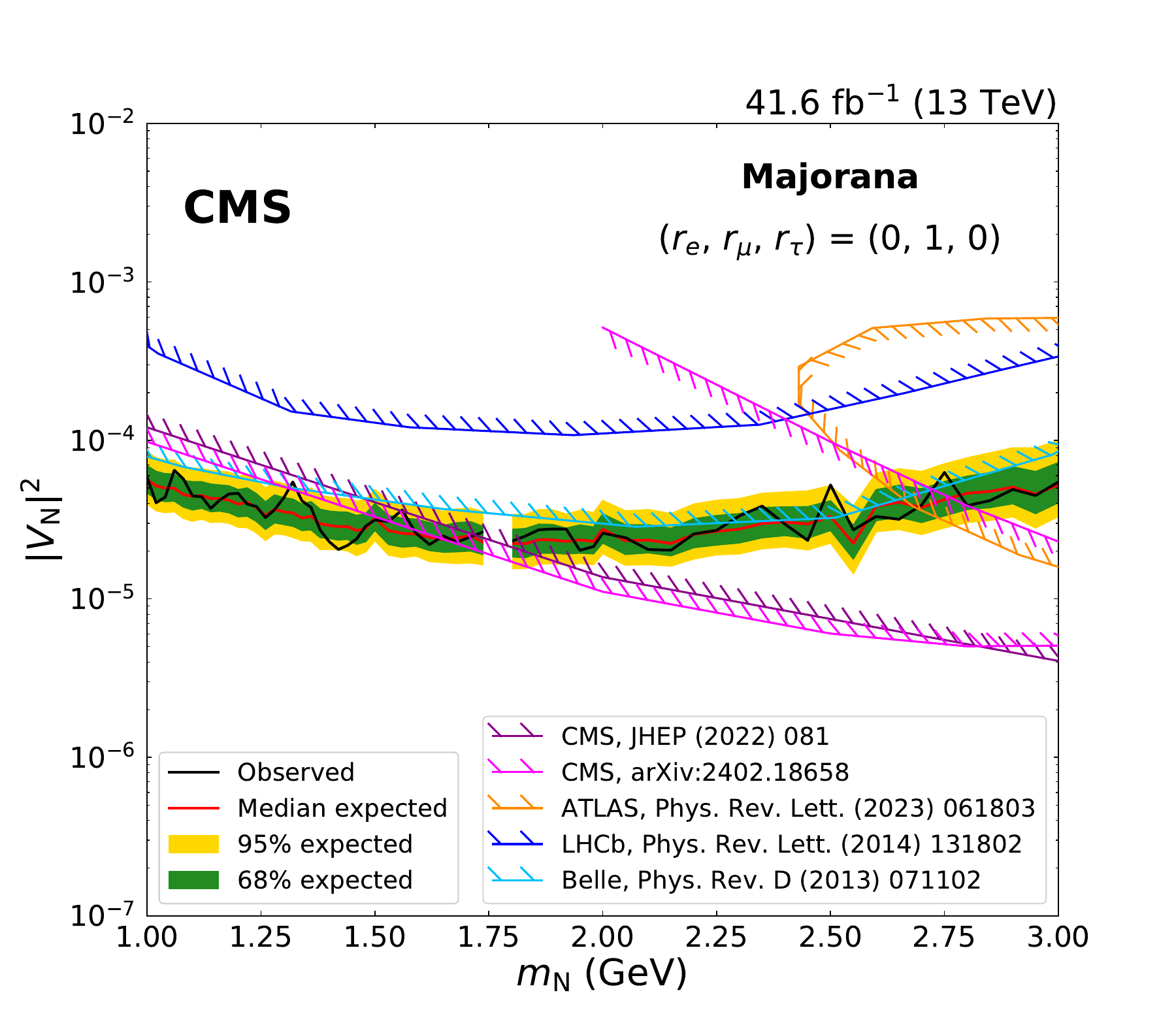}
\includegraphics[width=0.49\textwidth]{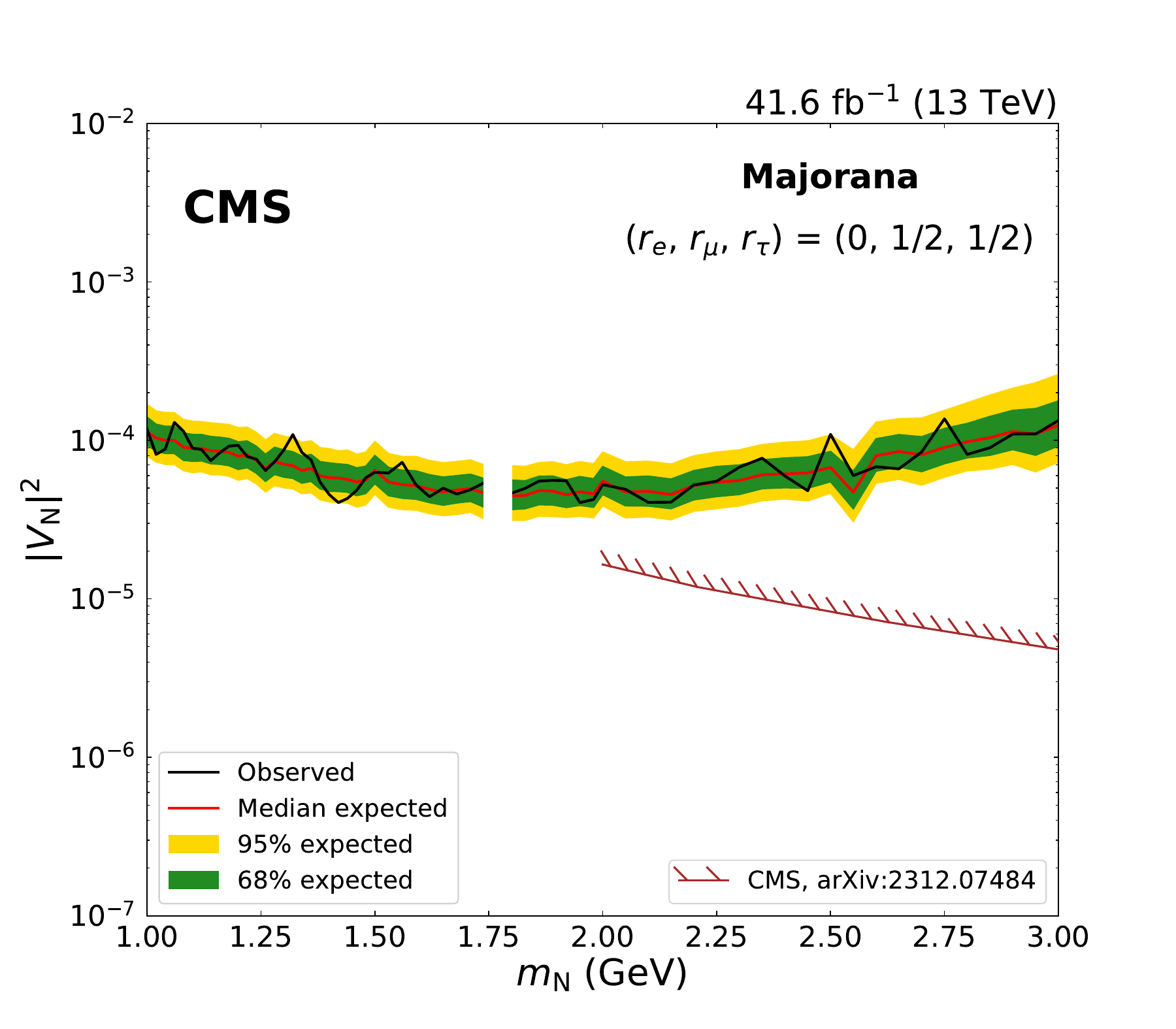}
\includegraphics[width=0.49\textwidth]{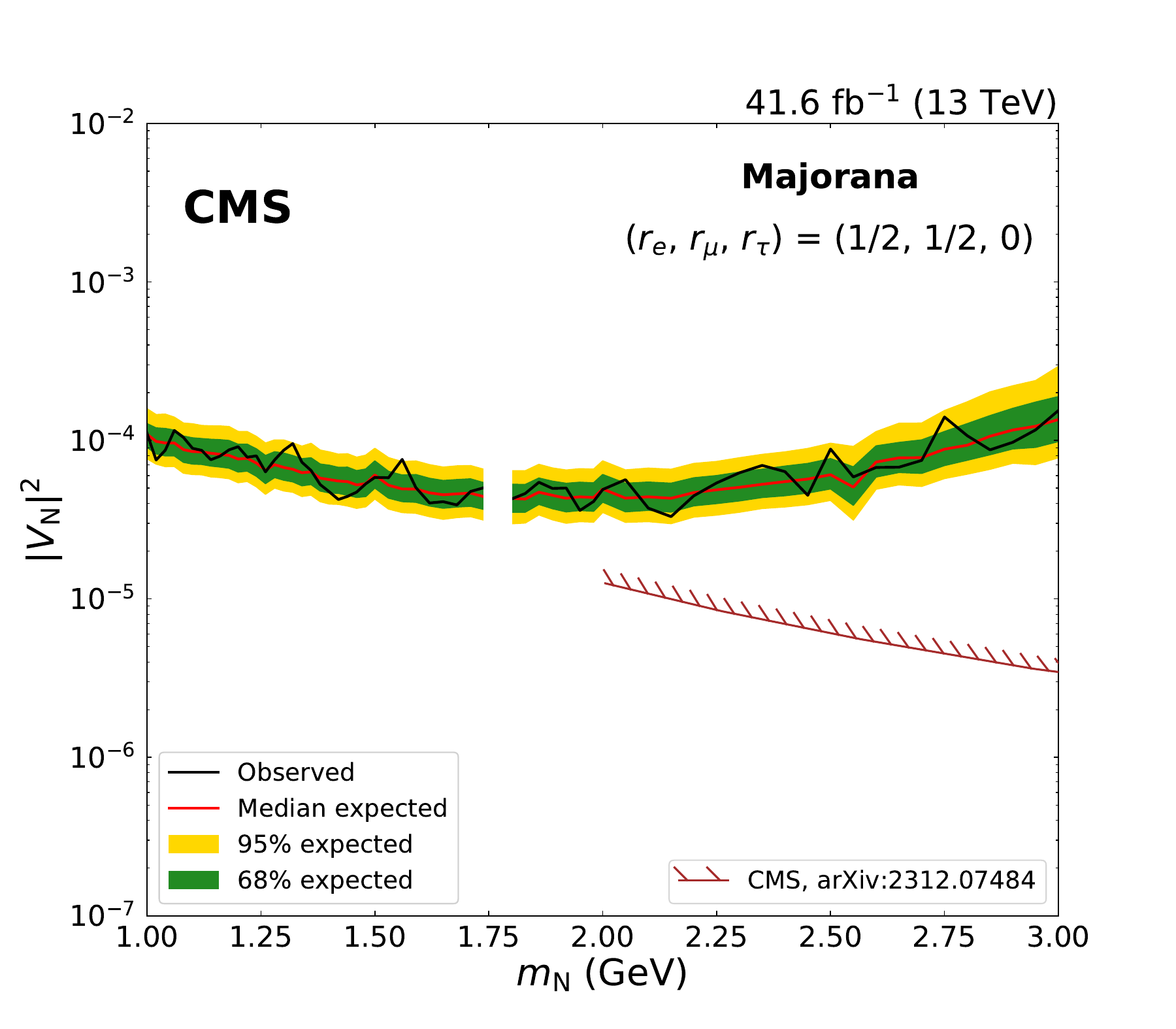}
\includegraphics[width=0.49\textwidth]{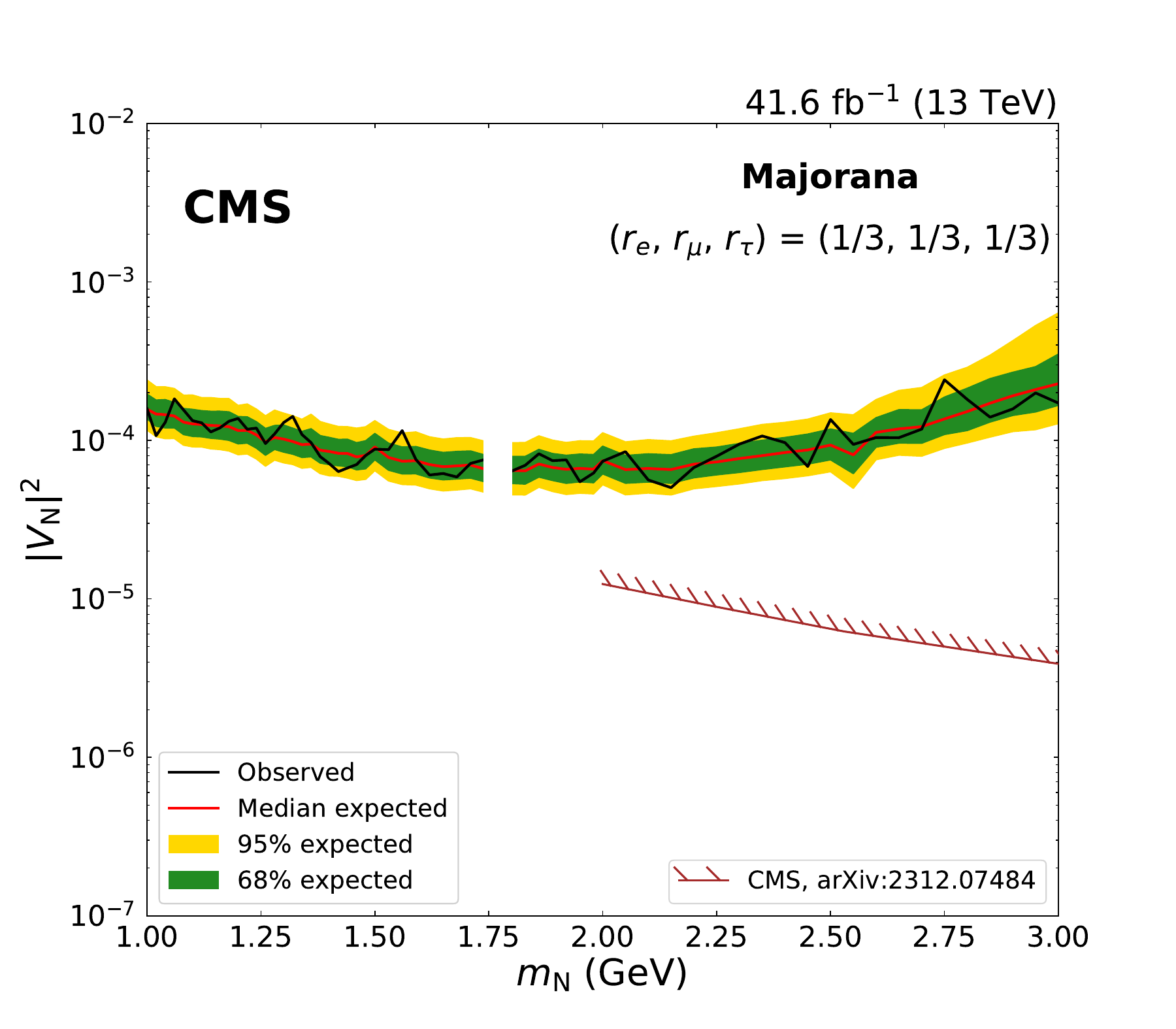}
\caption{Expected and observed 95\% \CL limits on \V as a function of \mN, in the Majorana scenario. On the upper row, the limits are derived uniquely with the dimuon channel, and are shown for the mixing scenarios (\fe, \fu, \ft) = (0, 1, 0) on the left and  for (\fe, \fu, \ft) = (0, 1/2, 1/2) on the right; on the lower row, the limits are obtained with the dimuon and mixed-flavour channel combined, for the mixing scenarios (\fe, \fu, \ft) = (1/2, 1/2, 0) on the left and  for (\fe, \fu, \ft) = (1/3, 1/3, 1/3) on the right. In the upper left figure, results from the CMS~\cite{WHNLpaper,CMS:2024ake}, ATLAS~\cite{ATLAS2022}, LHCb~\cite{LHCb:2014osd}, and Belle~\cite{Belle:2013ytx} Collaborations are shown as a comparison; in the other figures, results from the CMS Collaboration~\cite{EXO_21_013} are reported. The mass range with no results shown corresponds to the $\PDz$ meson veto listed in the lower part of Table~\ref{table_sm_resonances_veto}. 
\label{figure_limits_majorana}}
\end{figure}

\begin{figure}[tbh!]
\centering
\includegraphics[width=0.49\textwidth]{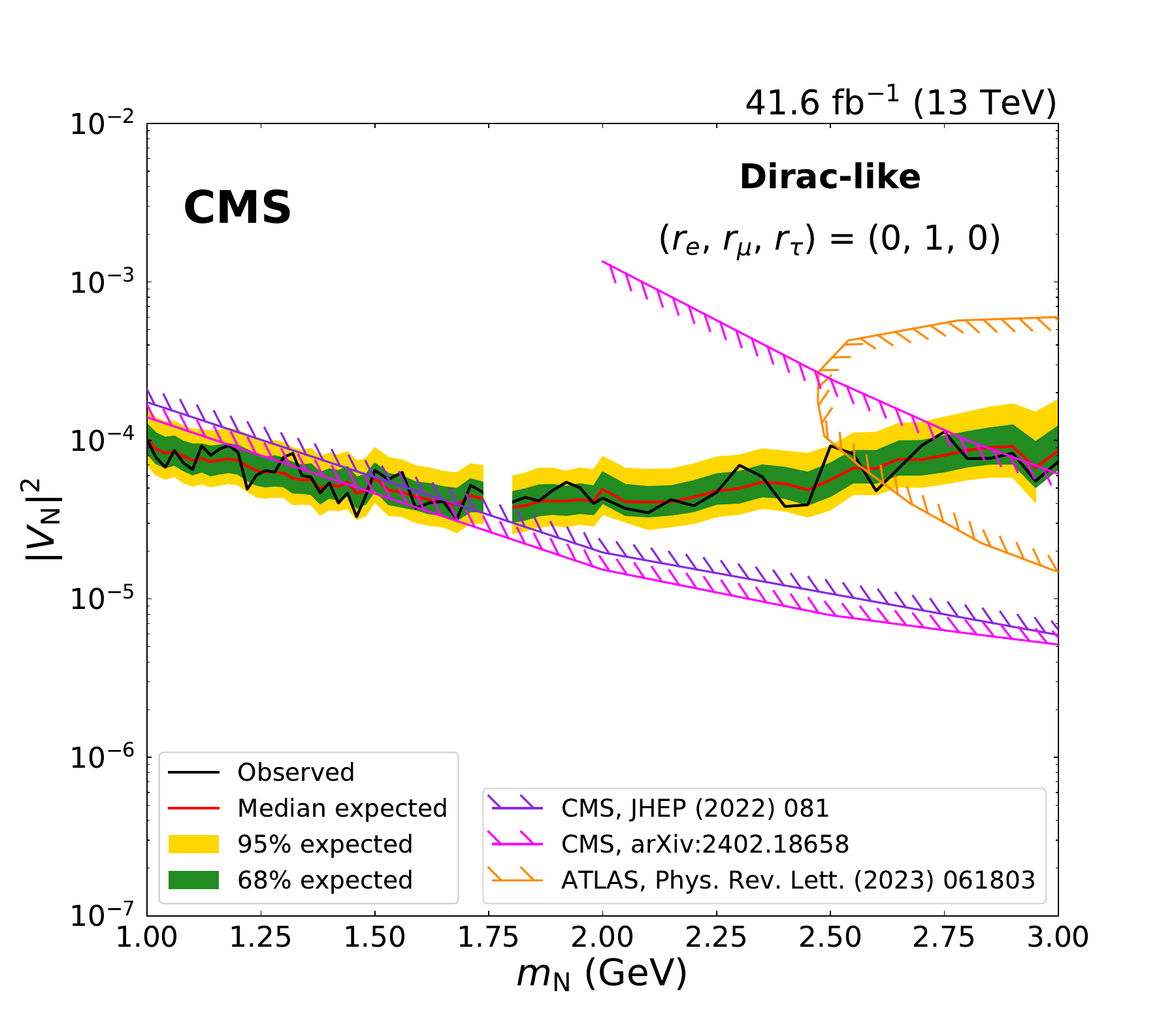}
\includegraphics[width=0.49\textwidth]{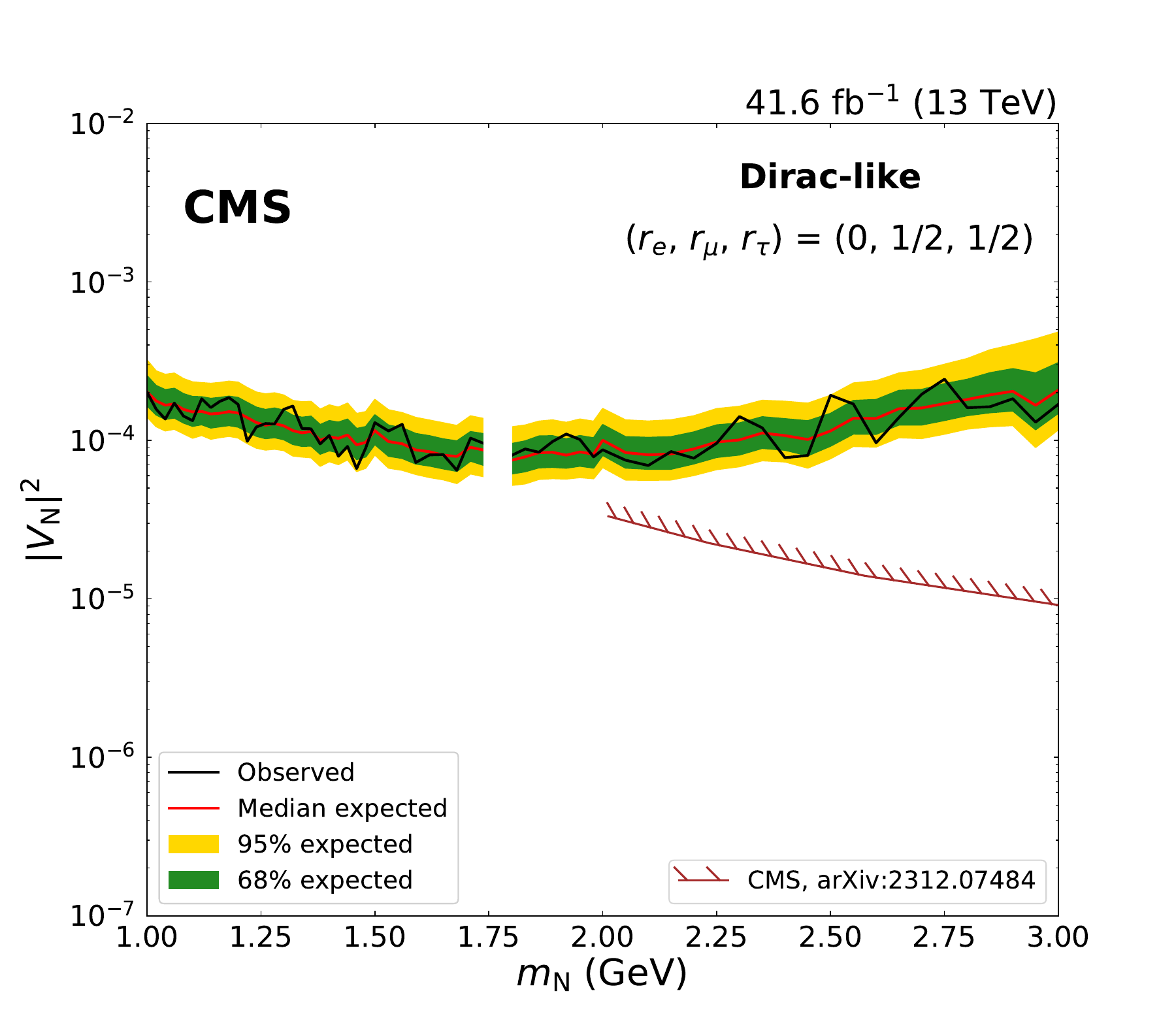}
\includegraphics[width=0.49\textwidth]{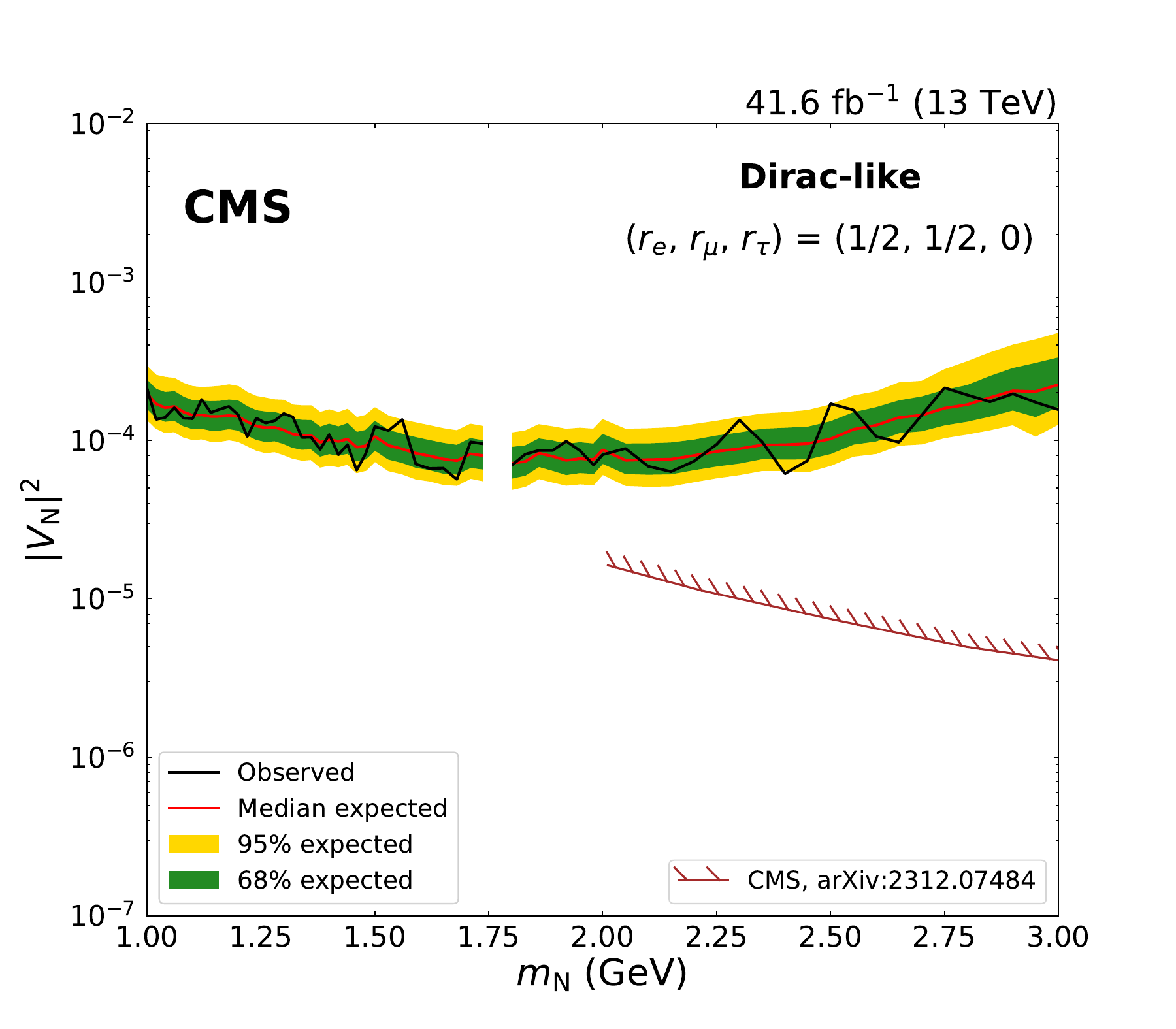}
\includegraphics[width=0.49\textwidth]{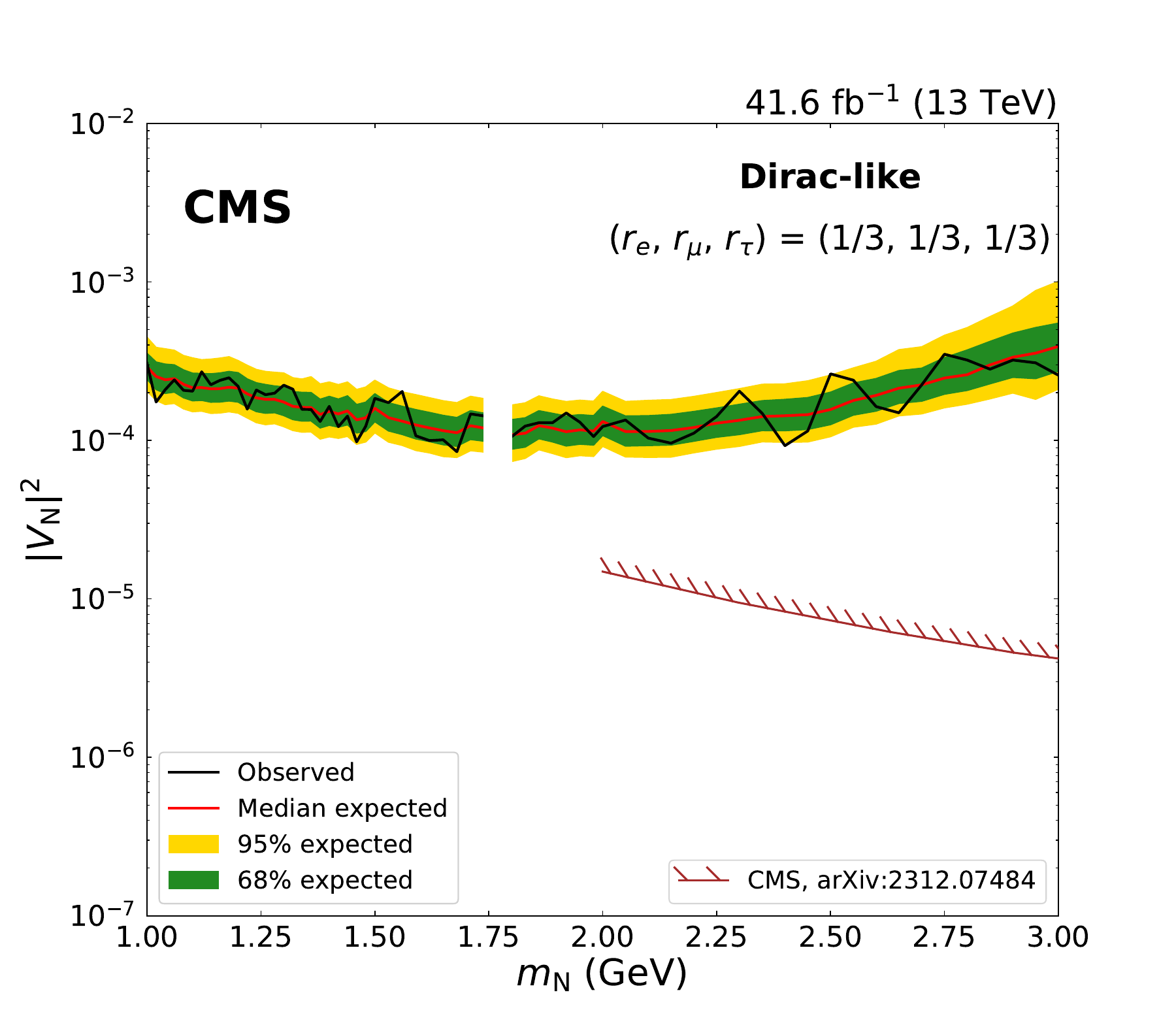}
\caption{Expected and observed 95\% \CL limits on \V as a function of \mN, in the Dirac-like scenario. On the upper row, the limits are derived uniquely with the dimuon channel, and are shown for the mixing scenarios (\fe, \fu, \ft) = (0, 1, 0) on the left and  for (\fe, \fu, \ft) = (0, 1/2, 1/2) on the right; on the lower row, the limits are obtained with the dimuon and mixed-flavour channel combined, for the mixing scenarios (\fe, \fu, \ft) = (1/2, 1/2, 0) on the left and  for (\fe, \fu, \ft) = (1/3, 1/3, 1/3) on the right. In the upper left figure, results from the CMS~\cite{WHNLpaper,CMS:2024ake} and ATLAS~\cite{ATLAS2022} Collaborations are shown as a comparison; in the other figures, results from the CMS Collaboration~\cite{EXO_21_013} are reported. The mass range with no results shown corresponds to the $\PDz$ meson veto listed in the lower part of Table~\ref{table_sm_resonances_veto}.
\label{figure_limits_dirac}}
\end{figure}

The upper exclusion limits on \V as a function of the \PN mass are shown in Figs.~\ref{figure_limits_majorana} and~\ref{figure_limits_dirac}, for the Majorana and Dirac-like cases, respectively. In each figure, four benchmark scenarios, chosen based on the proposal in Ref.~\cite{Drewes_2022}, are considered: two mixing scenarios derived with the dimuon channel only, with $(\fe,\,\fu,\,\ft) = (0,\,1,\,0)$ and $(\fe,\,\fu,\,\ft) = (0,\,1/2,\,1/2)$; and two mixing scenarios derived with the dimuon and mixed-flavour channels combined, namely $(\fe,\,\fu,\,\ft) = (1/2,\,1/2,\,0)$ and $(\fe,\,\fu,\,\ft) = (1/3,\,1/3,\,1/3)$. Using Eqs.~(\ref{equation_FN}) and~(\ref{eq:tau_vs_coupling}) and a reweighting procedure, we can obtain exclusion limits for an arbitrary mixing scenario.  

The best limit on \V is obtained for the muon exclusive mixing scenario $(\fe,\,\fu,\,\ft) = (0,\,1,\,0)$. This limit excludes at 95\% \CL values of $\V > 2.0\times 10^{-5}$ for a signal mass $\mN= 1.95\GeV$ for the Majorana case, and $\V > 3.2\times 10^{-5}$ for a signal mass $\mN= 1.68\GeV$ for the Dirac-like case. The best limits on \V for the other scenarios are summarized in Table~\ref{table_limits_v2}.

\begin{table}[tbp!]
\centering
\topcaption{Summary of the most stringent upper limits on \V at 95\% CL. For each scenario, the minimum excluded value of \V is reported together with the mass at which it occurs.\label{table_limits_v2}}
\begin{tabular}{lcccc}
\hline
(\fe, \fu, \ft) & Scenario & \V & Mass (GeV) \\
\hline
(0, 1, 0) & Majorana & $2.0\times 10^{-5}$ & 1.95\\
(0, 1/2, 1/2) & Majorana & $4.0\times 10^{-5}$ & 1.42  \\
(1/2, 1/2, 0) & Majorana & $3.3\times 10^{-5}$ & 2.15 \\
(1/3, 1/3, 1/3) & Majorana & $5.0\times 10^{-5}$ & 2.15 \\[\cmsTabSkip]
(0, 1, 0) & Dirac-like & $3.2\times 10^{-5}$ & 1.68 \\
(0, 1/2, 1/2) & Dirac-like & $6.5\times 10^{-5}$ & 1.68\\
(1/2, 1/2, 0) & Dirac-like & $5.7\times 10^{-5}$ & 1.68\\
(1/3, 1/3, 1/3) & Dirac-like & $8.5\times 10^{-5}$ & 1.68\\
\hline
\end{tabular}
\end{table}

The muon-exclusive mixing scenario allows a direct comparison with the results from previous work.
Compared to results of the ATLAS~\cite{ATLAS2022} and CMS~\cite{WHNLpaper} Collaborations, targeting long-lived HNLs produced in the decays of \PW~bosons, and analysing 139\fbinv and 138\fbinv of data, respectively, the limits are improved by up to a factor of about 3 for masses below 1.75\GeV for the Majorana case, and a factor of about 2 for masses below 1.7\GeV for the Dirac-like case. Furthermore, in the Majorana case, a direct comparison with results from the LHCb~\cite{LHCb:2014osd} (and revised in Ref.~\cite{LHCbPeskin}) and Belle~\cite{Belle:2013ytx} Collaborations, both targeting the same signal process as in the present search, is possible. Compared to the LHCb search, the limits are improved by up to one order of magnitude, while compared to the Belle results, the limits are improved by up to a factor of about 2 for all masses. 
Compared to other recent results from CMS~\cite{EXO_21_013,CMS:2024ake}, this analysis extends and improves the exclusion limits by up to a factor of about 2 in the mass region between 1 and 2\GeV.

The sensitivity for the three other mixing scenarios, with $\fu\neq 1$, is less than in the muon exclusive mixing scenario, as the reconstruction efficiency for the muon is greater than that of the electron with CMS. 

\begin{figure}[tbh!]
\centering
\includegraphics[width=0.49\textwidth]{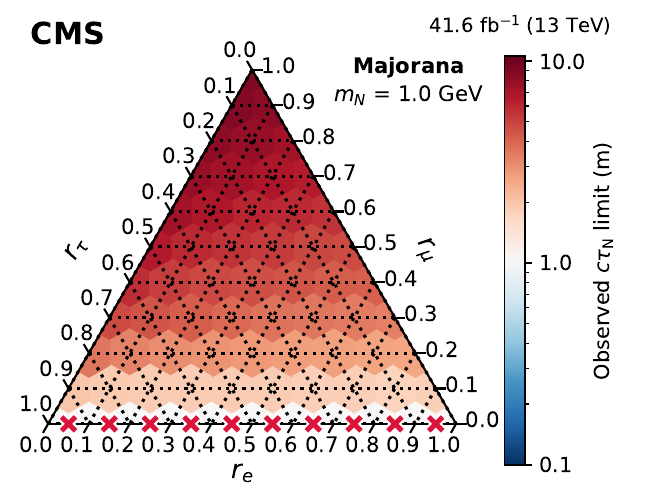}
\includegraphics[width=0.49\textwidth]{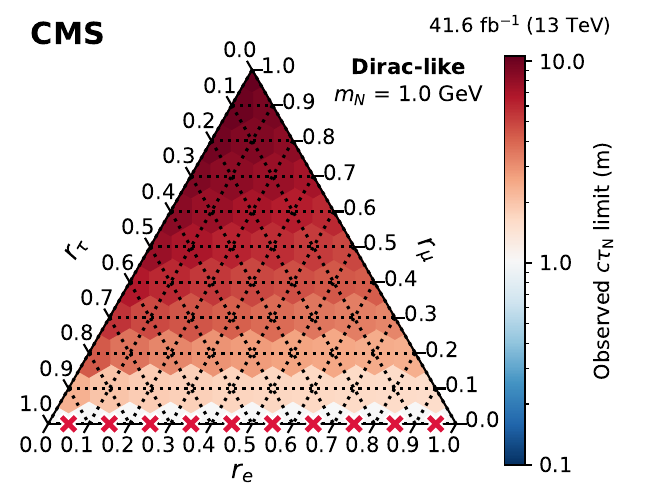}
\includegraphics[width=0.49\textwidth]{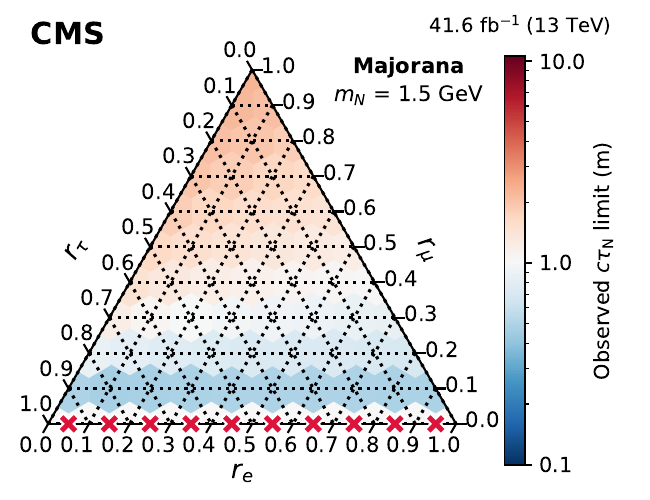}
\includegraphics[width=0.49\textwidth]{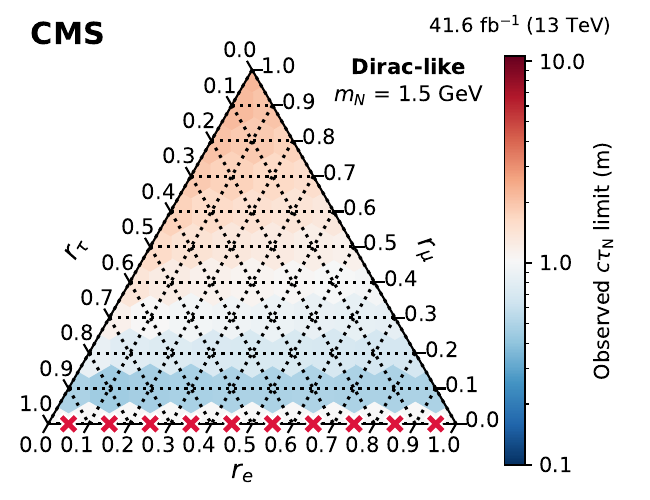}
\includegraphics[width=0.49\textwidth]{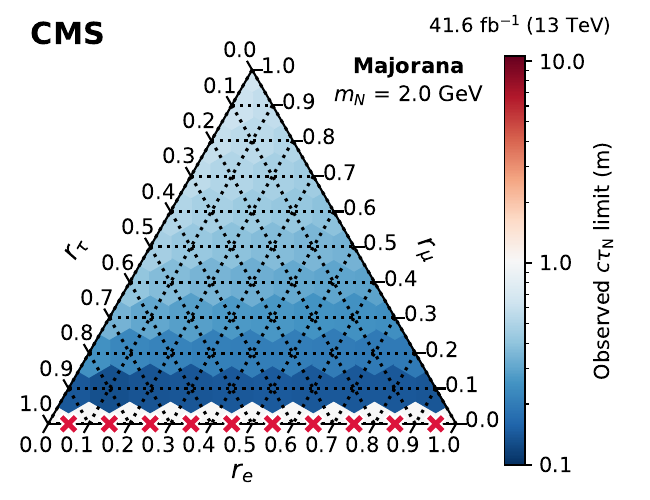}
\includegraphics[width=0.49\textwidth]{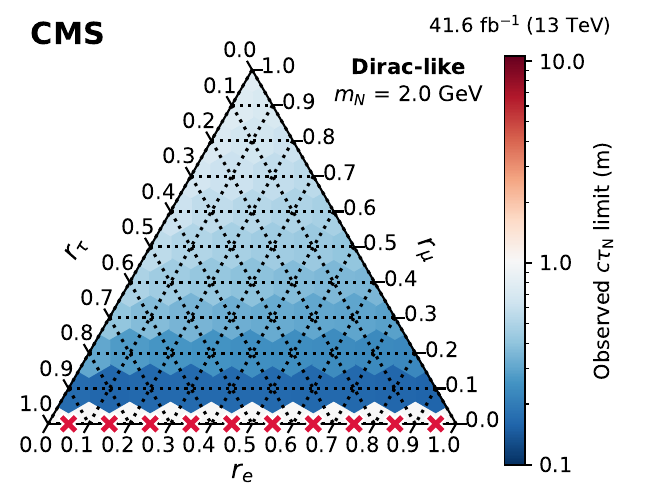}
\caption{Observed 95\% \CL lower limits on \ctau as functions of the mixing ratios (\fe, \fu, \ft) for fixed \PN masses of 1\GeV (upper row), 1.5\GeV (middle row), and 2\GeV (lower row), in the Majorana (left column) and Dirac-like (right column) scenarios. The red crosses indicate that there is no exclusion found for that point. The orientation of the value markers on each axis identifies the associated internal lines on the plot. \label{figure_ternary}}
\end{figure}

The observed lower limits at 95\% \CL on \ctau for 66 different mixing scenarios (\fe, \fu, \ft), for three values of masses, $\mN=1.0$, 1.5, and 2.0\GeV, are shown in Fig.~\ref{figure_ternary} for the Majorana and Dirac-like cases. The constraint $\fe + \fu + \ft = 1$ allows the values to be shown in the form of ternary plots. We verify that the limits improve with increasing values of \fu, while they slightly degrade with increasing values of \fe. The latter trend is explained by smaller values of $\widetilde{\Gamma}_\PGt$ with respect to $\widetilde{\Gamma}_\PGm$ and $\widetilde{\Gamma}_\Pe$ in Eq.~(\ref{eq:tau_vs_coupling}). The best limits on \ctau are obtained for a signal mass $\mN=1\GeV$ for the mixing scenario $(\fe,\,\fu,\,\ft) = (0,\,1,\,0)$. Values of $\ctau<9.2\unit{m}$ and $\ctau<10.5\unit{m}$ are excluded at 95\% \CL for the Majorana and Dirac-like cases, respectively. The most stringent limits on \ctau for the all masses are summarized in Table~\ref{table_limits_ctau} for the Majorana and Dirac-like hypotheses. Unlike the limits obtained on \V discussed above, the limits on \ctau are more stringent in the Dirac-like scenario than in the Majorana scenario. This behaviour is a consequence of $\widetilde{\Gamma}_\Pe$, $\widetilde{\Gamma}_\PGm$, and $\widetilde{\Gamma}_\PGt$ in Eq.~(\ref{eq:tau_vs_coupling}) being twice as low in the Dirac-like case as in the Majorana case.

\clearpage

\begin{table}[tbh!]
\centering
\topcaption{Summary of the most stringent lower limits on \ctau at 95\% CL, obtained for the mixing scenario $(\fe,\,\fu,\,\ft) = (0,\,1,\,0)$. The maximum excluded value of \ctau is reported for masses \mN = 1.0, 1.5 and 2.0, and for the Majorana and Dirac-like scenarios.\label{table_limits_ctau}}
\begin{tabular}{ccc}
\hline
Mass (GeV) & Scenario & \ctau (m) \\
\hline
1.0 & Majorana & 9.2 \\
1.5 & Majorana & 2.3 \\
2.0 & Majorana & 0.7 \\[\cmsTabSkip]
1.0 & Dirac-like & 10.5 \\
1.5 & Dirac-like & 2.2 \\
2.0 & Dirac-like & 0.8 \\
\hline
\end{tabular}
\end{table}

\section{Summary}\label{section_summary}

A search for long-lived heavy neutrinos, \PN, in the leptonic and semileptonic decays of \PB mesons produced in proton-proton collisions at $\sqrt{s}=13$\TeV has been performed. The search uses a special data sample, referred to as the B-parking data sample, accumulated by the CMS experiment during 2018. The sample corresponds to an integrated luminosity of 41.6\fbinv and contains of order $10^{10}$ \bbbar events. 

The search is based on the process $\PB\to\ell_\PB\PN \X$,  $\PN\to \ell^{\pm}\PGp^{\mp}$, where the charged leptons $\ell_\PB$ and $\ell$ are required to be $\ell_\PB\ell=\PGm\PGm,\,\PGm\Pe,\,\mathrm{or~}\Pe\PGm$; the hadronic recoil system, \X, is treated inclusively and is not reconstructed and the $\PB =(\PB_{{\PQu}}$, $\PB_{{\PQd}}$, $\PB_{{\PQs}}$, $\PB_{{\PQc}})$ decays are summed. 
Results are reported for the \PN mass range $1 < m_{\PN} < 3\GeV$.  

The main elements of the search signature are (i) two charged leptons, at least one of which must be a muon that satisfies the B-parking trigger requirements, (ii) a displaced vertex associated with the $\PN\to\ell^{\pm}\PGp^{\mp}$ decay, and (iii) a peak in the invariant mass distribution of the $ \ell^{\pm}\PGp^{\mp}$ system consistent with the expected signal shape. Backgrounds, which arise primarily from strong-interaction processes, are suppressed using a parametric neural network that considers a broad range of event properties. 

A search for \PN states is performed using simultaneous maximum likelihood fits to the $\ell^{\pm}\PGp^{\mp}$ invariant mass distributions in 24 mutually exclusive event categories. No significant excess of events over the SM background is observed in any of the fit regions.

The results are interpreted for the separate hypotheses of a Majorana or Dirac-like particle as (i) upper limits at 95\% \CL on $\abs{V_{\PN}}^2$ as functions of $m_{\PN}$, for representative scenarios specified by different values of the mixing ratios \fe, \fu, and \ft; and as (ii) lower limits at 95\% \CL on \ctau for 66 combinations of \fe, \fu, and \ft for signal masses $\mN =1.0$, 1.5, and 2.0\GeV. The most stringent limits are $\abs{V_{\PN}}^2 < 2.0\times 10^{-5}$ and $\ctau > 10.5\unit{m}$, obtained for the Majorana and Dirac-like cases, respectively, and for the scenario in which the \PN mixes exclusively with the muon sector. 

This search provides the most stringent exclusion limits on \V for masses $1 < \mN <1.7\GeV$ from a collider experiment to date. Assuming the benchmark scenario $(\fe,\,\fu,\,\ft) = (0,\,1,\,0)$ and the Majorana hypothesis, the exclusion is improved by almost one order of magnitude compared to LHCb~\cite{LHCb:2014osd}, and by up to a factor of about 2 compared to Belle~\cite{Belle:2013ytx} and the most stringent previous hadron collider result~\cite{CMS:2024ake}. Furthermore, the first upper limits on \V are set for the mass range $1 < \mN < 2\GeV$ for the mixing scenarios $(\fe,\,\fu,\,\ft) = (0,\,1/2,\,1/2),\, (1/2,\,1/2,\,0)$, and $(1/3,\,1/3,\,1/3)$. Finally, lower limits on \ctau in the form of ternary plots for masses $\mN \leq 2.0\GeV$ are presented for the first time.

\begin{acknowledgments}
We congratulate our colleagues in the CERN accelerator departments for the excellent performance of the LHC and thank the technical and administrative staffs at CERN and at other CMS institutes for their contributions to the success of the CMS effort. In addition, we gratefully acknowledge the computing centres and personnel of the Worldwide LHC Computing Grid and other centres for delivering so effectively the computing infrastructure essential to our analyses. Finally, we acknowledge the enduring support for the construction and operation of the LHC, the CMS detector, and the supporting computing infrastructure provided by the following funding agencies: SC (Armenia), BMBWF and FWF (Austria); FNRS and FWO (Belgium); CNPq, CAPES, FAPERJ, FAPERGS, and FAPESP (Brazil); MES and BNSF (Bulgaria); CERN; CAS, MoST, and NSFC (China); MINCIENCIAS (Colombia); MSES and CSF (Croatia); RIF (Cyprus); SENESCYT (Ecuador); ERC PRG, RVTT3 and MoER TK202 (Estonia); Academy of Finland, MEC, and HIP (Finland); CEA and CNRS/IN2P3 (France); SRNSF (Georgia); BMBF, DFG, and HGF (Germany); GSRI (Greece); NKFIH (Hungary); DAE and DST (India); IPM (Iran); SFI (Ireland); INFN (Italy); MSIP and NRF (Republic of Korea); MES (Latvia); LMTLT (Lithuania); MOE and UM (Malaysia); BUAP, CINVESTAV, CONACYT, LNS, SEP, and UASLP-FAI (Mexico); MOS (Montenegro); MBIE (New Zealand); PAEC (Pakistan); MES and NSC (Poland); FCT (Portugal); MESTD (Serbia); MCIN/AEI and PCTI (Spain); MOSTR (Sri Lanka); Swiss Funding Agencies (Switzerland); MST (Taipei); MHESI and NSTDA (Thailand); TUBITAK and TENMAK (Turkey); NASU (Ukraine); STFC (United Kingdom); DOE and NSF (USA).

\hyphenation{Rachada-pisek} Individuals have received support from the Marie-Curie programme and the European Research Council and Horizon 2020 Grant, contract Nos.\ 675440, 724704, 752730, 758316, 765710, 824093, 101115353, and COST Action CA16108 (European Union); the Leventis Foundation; the Alfred P.\ Sloan Foundation; the Alexander von Humboldt Foundation; the Science Committee, project no. 22rl-037 (Armenia); the Belgian Federal Science Policy Office; the Fonds pour la Formation \`a la Recherche dans l'Industrie et dans l'Agriculture (FRIA-Belgium); the Agentschap voor Innovatie door Wetenschap en Technologie (IWT-Belgium); the F.R.S.-FNRS and FWO (Belgium) under the ``Excellence of Science -- EOS" -- be.h project n.\ 30820817; the Beijing Municipal Science \& Technology Commission, No. Z191100007219010 and Fundamental Research Funds for the Central Universities (China); the Ministry of Education, Youth and Sports (MEYS) of the Czech Republic; the Shota Rustaveli National Science Foundation, grant FR-22-985 (Georgia); the Deutsche Forschungsgemeinschaft (DFG), under Germany's Excellence Strategy -- EXC 2121 ``Quantum Universe" -- 390833306, and under project number 400140256 - GRK2497; the Hellenic Foundation for Research and Innovation (HFRI), Project Number 2288 (Greece); the Hungarian Academy of Sciences, the New National Excellence Program - \'UNKP, the NKFIH research grants K 124845, K 124850, K 128713, K 128786, K 129058, K 131991, K 133046, K 138136, K 143460, K 143477, 2020-2.2.1-ED-2021-00181, and TKP2021-NKTA-64 (Hungary); the Council of Science and Industrial Research, India; ICSC -- National Research Centre for High Performance Computing, Big Data and Quantum Computing, funded by the EU NexGeneration program (Italy); the Latvian Council of Science; the Ministry of Education and Science, project no. 2022/WK/14, and the National Science Center, contracts Opus 2021/41/B/ST2/01369 and 2021/43/B/ST2/01552 (Poland); the Funda\c{c}\~ao para a Ci\^encia e a Tecnologia, grant CEECIND/01334/2018 (Portugal); the National Priorities Research Program by Qatar National Research Fund; MCIN/AEI/10.13039/501100011033, ERDF ``a way of making Europe", and the Programa Estatal de Fomento de la Investigaci{\'o}n Cient{\'i}fica y T{\'e}cnica de Excelencia Mar\'{\i}a de Maeztu, grant MDM-2017-0765 and Programa Severo Ochoa del Principado de Asturias (Spain); the Chulalongkorn Academic into Its 2nd Century Project Advancement Project, and the National Science, Research and Innovation Fund via the Program Management Unit for Human Resources \& Institutional Development, Research and Innovation, grant B37G660013 (Thailand); the Kavli Foundation; the Nvidia Corporation; the SuperMicro Corporation; the Welch Foundation, contract C-1845; and the Weston Havens Foundation (USA).    
\end{acknowledgments}

\bibliography{auto_generated} 

\cleardoublepage \appendix\section{The CMS Collaboration \label{app:collab}}\begin{sloppypar}\hyphenpenalty=5000\widowpenalty=500\clubpenalty=5000\input{EXO-22-019-public-authorlist.tex}\end{sloppypar}
\end{document}

%% file: EXO-22-019-public-authorlist.tex
\cmsinstitute{Yerevan Physics Institute, Yerevan, Armenia}
{\tolerance=6000
A.~Hayrapetyan, A.~Tumasyan\cmsAuthorMark{1}\cmsorcid{0009-0000-0684-6742}
\par}
\cmsinstitute{Institut f\"{u}r Hochenergiephysik, Vienna, Austria}
{\tolerance=6000
W.~Adam\cmsorcid{0000-0001-9099-4341}, J.W.~Andrejkovic, T.~Bergauer\cmsorcid{0000-0002-5786-0293}, S.~Chatterjee\cmsorcid{0000-0003-2660-0349}, K.~Damanakis\cmsorcid{0000-0001-5389-2872}, M.~Dragicevic\cmsorcid{0000-0003-1967-6783}, P.S.~Hussain\cmsorcid{0000-0002-4825-5278}, M.~Jeitler\cmsAuthorMark{2}\cmsorcid{0000-0002-5141-9560}, N.~Krammer\cmsorcid{0000-0002-0548-0985}, A.~Li\cmsorcid{0000-0002-4547-116X}, D.~Liko\cmsorcid{0000-0002-3380-473X}, I.~Mikulec\cmsorcid{0000-0003-0385-2746}, J.~Schieck\cmsAuthorMark{2}\cmsorcid{0000-0002-1058-8093}, R.~Sch\"{o}fbeck\cmsorcid{0000-0002-2332-8784}, D.~Schwarz\cmsorcid{0000-0002-3821-7331}, M.~Sonawane\cmsorcid{0000-0003-0510-7010}, S.~Templ\cmsorcid{0000-0003-3137-5692}, W.~Waltenberger\cmsorcid{0000-0002-6215-7228}, C.-E.~Wulz\cmsAuthorMark{2}\cmsorcid{0000-0001-9226-5812}
% ...
\par}
\vskip\cmsinstskip
\dag:~Deceased\\
$^{1}$Also at Yerevan State University, Yerevan, Armenia\\
$^{2}$Also at TU Wien, Vienna, Austria\\
% ...
